ΤΜΗΜΑ ΜΗΧΑΝΙΚΩΝ ΗΛΕΚΤΡΟΝΙΚΩΝ ΥΠΟΛΟΓΙΣΤΩΝ ΚΑΙ ΠΛΗΡΟΦΟΡΙΚΗΣ

*Μεταπτυχιακή Διπλωματική Εργασία*

# Ανάπτυξη Συστήματος Συστάσεων Συνεργατικής Διήθησης με χρήση Ιεραρχικών Αλγορίθμων Κατάταξης

της

## Μαριάννας Κουνέλη

Επιβλέπων :
Καθηγητής Γιάννης Γαροφαλάκης

Πανεπιστήμιο Πατρών

Τμήμα Μηχανικών Ηλεκτρονικών Υπολογιστών και Πληροφορικής

*Μεταπτυχιακή Διπλωματική Εργασία*

# Ανάπτυξη Συστήματος Συστάσεων Συνεργατικής Διήθησης με χρήση Ιεραρχικών Αλγορίθμων Κατάταξης

## Κουνέλη Μαριάννα, Α.Μ 687

| Καθηγητής | Επίκουρος Καθηγητής | Αναπληρωτής Καθηγητής |
|---|---|---|
| Γιάννης Γαροφαλάκης | Χρήστος Μακρής | Ιωάννης Χατζηλυγερούδης |
| ............................. | ............................. | ............................. |
| ............................. | ............................. | ............................. |
| ............................. | ............................. | ............................. |

Στην οικογένειά μου που έχει κάνει
τα πάντα για μένα...

# Ευχαριστίες

Θα ήθελα να απευθύνω τις ιδιαίτερες ευχαριστίες μου σε όλους όσους συνέβαλλαν, με τον ένα ή τον άλλο τρόπο, στην εκπόνηση της μεταπτυχιακής μου διπλωματικής εργασίας.

Καταρχήν, θα ήθελα να ευχαριστήσω τον καθηγητή μου και επιβλέποντα, κ. Γιάννη Γαροφαλάκη για την πολύτιμη καθοδήγηση και βοήθεια.

Επίσης, θα ήθελα να ευχαριστήσω τους καθηγητές κ. Χρήστο Μακρή και κ. Ιωάννη Χατζηλυγερούδη που με εμπιστεύτηκαν και δέχτηκαν να συμμετάσχουν στην τριμελή επιτροπή αξιολόγησης της διπλωματικής μου.

Ιδιαίτερες ευχαριστίες θα ήθελα να απευθύνω στον υποψήφιο διδάκτορα Αθανάσιο Ν.Νικολακόπουλο για την πολύτιμη συμβολή του, τόσο καθ'όλη την διάρκεια του πειραματικού μέρους όσο και κατά τη διαδικασία της συγγραφής. Χωρίς την απλόχερη βοήθειά του και την επιστημονική του συνδρομή, η ολοκλήρωση της διπλωματικής θα ήταν αδύνατη. Οι ατέλειωτες ώρες, οι υποδείξεις, η προθυμία του καθώς και η υπομονή του, οδήγησαν στην ομαλή διεκπεραίωσή της.

Τέλος, οφείλω να ευχαριστήσω θερμά την οικογένειά μου και τον Παναγιώτη που είναι δίπλα μου, για τη στήριξη, τη συμπαράσταση και την κατανόησή τους σε όλη την διάρκεια των μεταπτυχιακών μου σπουδών.

# Περίληψη

Σκοπός της παρούσας διπλωματικής διατριβής είναι η μελέτη και ανάπτυξη ενός νέου αλγοριθμικού πλαισίου *Συνεργατικής Διήθησης* (CF) για την παραγωγή συστάσεων. Η μέθοδος που προτείνουμε, βασίζεται στην εκμετάλλευση της ιεραρχικής διάρθρωσης του χώρου αντικειμένων και πατά διαισθητικά στην ιδιότητα της "*Σχεδόν Πλήρους Αναλυσιμότητας*" (NCD) η οποία είναι συνυφασμένη με τη δομή της πλειοψηφίας των ιεραρχικών συστημάτων.

Η Συνεργατική Διήθηση αποτελεί ίσως την πιο πετυχημένη οικογένεια τεχνικών για την παραγωγή συστάσεων. Η μεγάλη απήχησή της στο διαδίκτυο αλλά και η ευρεία εφαρμογή της σε σημαντικά εμπορικά περιβάλλοντα, έχουν οδηγήσει στη σημαντική ανάπτυξη της θεωρίας την τελευταία δεκαετία, όπου μια ευρεία ποικιλία αλγορίθμων και μεθόδων έχουν προταθεί. Ωστόσο, παρά την πρωτοφανή τους επιτυχία οι CF μέθοδοι παρουσιάζουν κάποιους σημαντικούς περιορισμούς συμπεριλαμβανομένης της *επεκτασιμότητας* και της *αραιότητας* των δεδομένων. Τα προβλήματα αυτά επιδρούν αρνητικά στην ποιότητα των παραγόμενων συστάσεων και διακυβεύουν την εφαρμοσιμότητα πολλών CF αλγορίθμων σε ρεαλιστικά σενάρια.

Χτίζοντας πάνω στη διαίσθηση πίσω από τον αλγόριθμο *NCDawareRank* - μίας γενικής μεθόδου υπολογισμού διανυσμάτων κατάταξης ιεραρχικά δομημένων γράφων - και της σχετικής με αυτόν έννοιας της *NCD εγγύτητας*, προβαίνουμε σε μία μοντελοποίηση του συστήματος με τρόπο που φωτίζει τα ενδημικά του χαρακτηριστικά και προτείνουμε έναν νέο αλγοριθμικό πλαίσιο συστάσεων, τον **HIR**. Στο επίκεντρο της προσέγγισής μας είναι η προσπάθεια να συνδυάσουμε τις άμεσες με τις NCD, "γειτονιές" των αντικειμένων ώστε να πετύχουμε μεγαλύτερης ακρίβειας χαρακτηρισμό των πραγματικών συσχετισμών μεταξύ των στοιχείων του χώρου αντικειμένων, με σκοπό την βελτίωση της ποιότητας των συστάσεων αλλά και την αντιμετώπιση της εγγενούς αραιότητας και των προβλημάτων που αυτή συνεπάγεται.

Για να αξιολογήσουμε την απόδοση της μεθόδου μας υλοποιούμε και εφαρμόζουμε τον HIR στο κλασικό *movie recommendation* πρόβλημα και παραθέτουμε μια σειρά από πειράματα χρησιμοποιώντας το `MovieLens Dataset`. Τα πειράματά μας δείχνουν πως ο HIR με την εκμετάλλευση της ιδέας της NCD εγγύτητας καταφέρνει να πετύχει λίστες συστάσεων υψηλότερης ποιότητας σε σύγκριση με τις άλλες state-of-the-art μεθόδους που έχουν προταθεί στη βιβλιογραφία, σε ευρέως χρησιμοποιούμενες μετρικές (micro- και macro-DOA), αποδεικνύοντας την ίδια στιγμή πως είναι λιγότερο επιρρεπής στα προβλήματα που σχετίζονται με την αραιότητα και έχοντας παράλληλα ανταγωνιστικό προφίλ πολυπλοκότητας και απαιτήσεις αποθήκευσης.

**Λέξεις Κλειδιά**

Συστήματα συστάσεων, Συνεργατική Διήθηση, Αραιότητα, Σχεδόν Πλήρους Αναλυσιμότητα, Αλγόριθμοι Κατάταξης, Πειράματα

# Abstract

The purpose of this master's thesis is to study and develop a new algorithmic framework for collaborative filtering (CF) to generate recommendations. The method we propose is based on the exploitation of the hierarchical structure of the item space and intuitively "stands" on the property of *Near Complete Decomposability* (NCD) which is inherent in the structure of the majority of hierarchical systems.

Collaborative Filtering is one of the most successful families of recommendations methods. The great impact of CF on Web applications, and its wide deployment in important commercial environments, have led to the significant development of the theory, with a wide variety of algorithms and methods being proposed. However, despite their unprecedented success, CF methods present some important limitations including scalability and data sparsity. These problems have a negative impact of the quality of the recommendations and jeopardize the applicability of many CF algorithms in realistic scenarios.

Building on the intuition behind the NCDawareRank algorithm and its related concept of *NCD proximity*, we model our system in a way that illuminates its endemic characteristics and we propose a new algorithmic framework for recommendations, called HIR. We focus on combining the direct with the NCD " neighborhoods" of items to achieve better characterization of the inter-item relations, in order to improve the quality of recommendations and alleviate sparsity related problems.

To evaluate the merits of our method, we implement and apply HIR in the classic movie recommendation problem, running a number of experiments on the standard MovieLens dataset. Our experiments show that HIR manages to create recommendation lists with higher quality compared with other state-of-the-art methods proposed in the literature, in widely used metrics (micro- and macro- DOA), demonstrating at the same time that it is less prone to low density related problems being at the same time very efficient in both complexity and storage requirements.

# Περιεχόμενα

# Κατάλογος σχημάτων

# ΚΑΤΑΛΟΓΟΣ ΣΧΗΜΑΤΩΝ

# Κατάλογος πινάκων

# ΚΑΤΑΛΟΓΟΣ ΠΙΝΑΚΩΝ

# Listings

# LISTINGS

# Μέρος I

# Συστήματα Συστάσεων

# Κεφάλαιο 1

# Εισαγωγή

*What is the best holiday for me and my family?*
*Which book should I buy for my next vacation?*
*Which web sites will I find interesting?*
*Which digital camera should I buy?*
*Which movie should I rent?*

## 1.1 Συστήματα Συστάσεων

Τα *Συστήματα Συστάσεων (Recommender Systems, RS)* είναι εργαλεία λογισμικού, τα οποία χρησιμοποιούνται από τα περισσότερα ηλεκτρονικά εμπορικά καταστήματα προκειμένου να κάνουν έξυπνες και γρήγορες εξατομικευμένες συστάσεις στους χρήστες. Η ανάπτυξή τους ξεκίνησε από μια απλή παρατήρηση: τα άτομα συχνά βασίζονται στις συστάσεις που τους παρέχονται από άλλους, στη λήψη καθημερινών αποφάσεών τους. Για παράδειγμα, είναι σύνηθες άτομα τα οποία θέλουν να διαβάσουν ένα βιβλίο, να βασίζονται στις απόψεις ατόμων που το έχουν ήδη διαβάσει, οι εργοδότες κατά τη διαδικασία πρόσληψης να υπολογίζουν σε επιστολές για την απόφασή τους, και τέλος τα άτομα τα οποία επιθυμούν να δουν μια ταινία, να διαβάζουν και να βασίζονται στις κριτικές που έχει πάρει η ταινία αυτή.

Τα συστήματα αυτά μπορούν να ιδωθούν σαν μέθοδοι φιλτραρίσματος πληροφορίας και απευθύνονται κυρίως σε άτομα που δε διαθέτουν επαρκή προσωπική εμπειρία ή την ικανότητα να αξιολογήσουν τον συντριπτικό αριθμό των διαφορετικών αντικειμένων που υπάρχουν στο διαδίκτυο. Ένα παράδειγμα ενός συστήματος συστάσεων παρουσιάζεται στο Σχήμα 1.1.

Προκειμένου να παράξουν εξατομικευμένες συστάσεις, τα RS συλλέγουν από τους χρήστες τις προτιμήσεις τους, οι οποίες είτε *εκφράζονται ρητά*, όπως είναι οι βαθμολογίες που δίνουν για τα αντικείμενα, ή *συμπεραίνονται έμμεσα* από την αλληλεπίδραση του χρήστη

# 1. ΕΙΣΑΓΩΓΗ

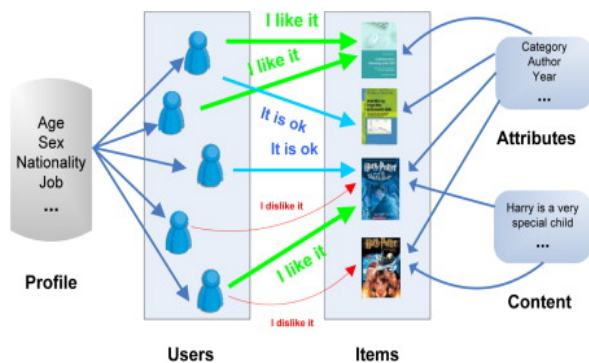

Σχήμα 1.1: Η εικόνα αυτή δείχνει ένα *σύστημα συστάσεων* το οποίο αποτελείται από πέντε χρήστες και τέσσερα βιβλία. Οι βασικές πληροφορίες που περιέχονται από κάθε σύστημα συστάσεων είναι οι σχέσεις μεταξύ των χρηστών και των αντικειμένων που μπορεί να εκπροσωπούνται από ένα διμερές γράφο. Επιπλέον, παρουσιάζει κάποιες πρόσθετες πληροφορίες που συχνά αξιοποιούνται κατά τη διάρκεια σχεδιασμού των αλγορίθμων σύστασης, όπως είναι το προφίλ των χρηστών, οι ιδιότητες που έχουν τα αντικείμενα καθώς και το περιεχόμενό τους [48].

με το σύστημα. Για παράδειγμα, ένα RS μπορεί να θεωρήσει την πλοήγηση ενός χρήστη σε μια συγκεκριμένη σελίδα ως μια έμμεση ένδειξη της προτίμησής του προς τα αντικείμενα που εμφανίζονται σε αυτή τη σελίδα

Για την παραγωγή συστάσεων έχουν αναπτυχθεί διάφορες τεχνολογίες. Την πιο επιτυχημένη αποτελεί η *Συνεργατική Διήθηση (Collaborative Filtering, CF)*· η μεγάλη απήχησή της στο διαδίκτυο αλλά και η ευρεία εφαρμογή της σε σημαντικά εμπορικά περιβάλλοντα, έχουν οδηγήσει στη σημαντική ανάπτυξη της θεωρίας την τελευταία δεκαετία, όπου μια ευρεία ποικιλία αλγορίθμων και μεθόδων έχουν προταθεί [78]. Το σκεπτικό πίσω από τις CF μεθόδους είναι πως αν ο ενεργός χρήστης[1] συμφώνησε στο παρελθόν με κάποιους χρήστες, τότε οι άλλες συστάσεις που προέρχονται από αυτούς θα πρέπει να είναι σχετικές, και κοντά στα ενδιαφέροντά του.

Άλλες σημαντικές τεχνολογίες συστημάτων συστάσεων, αποτελούν η *διήθηση με βάση το περιεχόμενο (content-based filtering)* και η *υβριδική (hybrid)* προσέγγιση. Στην περίπτωση της content-based διήθησης, το σύστημα μαθαίνει να προτείνει αντικείμενα σε ένα χρήστη με βάση το τι έχει αγοράσει στο παρελθόν. Συγκεκριμένα, αναλύει το περιεχόμενο των ήδη βαθμολογημένων - από αυτόν - αντικειμένων και χτίζει το προφίλ του. Κατά τη διαδικασία παραγωγής συστάσεων, το content-based σύστημα ταιριάζει τα χαρακτηριστικά - στα οποία συμπεριλαμβάνονται οι προτιμήσεις του - που υπάρχουν στο προφίλ του χρήστη με τα χαρακτηριστικά του περιεχομένου των αντικειμένων που δεν έχει αλληλεπιδράσει ακόμα.

Στην υβριδική προσέγγιση, τα συστήματα βασίζονται στο συνδυασμό των διαφόρων τε-χνικών των RS. Ένα υβριδικό σύστημα συνδυάζει την content-based και την CF ώστε να χρησιμοποιήσει τα πλεονεκτήματα της μιας προκειμένου να διορθώσει τα μειονεκτήματα της άλλης. Για παράδειγμα, οι μέθοδοι συνεργατικής διήθησης υποφέρουν από το πρόβλημα των "νέων αντικειμένων", δηλαδή οι μέθοδοι αυτοί δεν μπορούν να προτείνουν νέα αντικείμενα στους χρήστες αφού έχουν βαθμολογηθεί λίγο ή και καθόλου. Αυτό δεν θεωρείται περιορισμός για τις content-based προσεγγίσεις εφόσον η πρόβλεψη για νέα αντικείμενα βασίζεται στα χαρακτηριστικά τους τα οποία είναι τις περισσότερες φορές εύκολα διαθέσιμα. Από την άλλη μεριά, οι CF τεχνικές μπορούν να προτείνουν στους χρήστες αντικείμενα με πολύ διαφορετικό περιεχόμενο - το μόνο που αρκεί είναι να έχουν δείξει παρόμοιοι χρήστες ενδιαφέρον γι αυτά τα αντικείμενα - κάτι που δεν ισχύει στις content-based.

---
[1] σαν ενεργός χρήστης θεωρείται αυτός που ψάχνει για συστάσεις

Σκοπός της διατριβής αυτής είναι η μελέτη και η ανάπτυξη ενός **νέου αλγοριθμικού πλαισίου** για την παραγωγή συστάσεων, που βασίζεται στη Συνεργατική Διήθηση. Η μέθοδος που προτείνουμε, εκμεταλλεύεται την **ιεραρχική διάρθρωση του χώρου αντικειμένων** και πατά διαισθητικά στην ιδιότητα της "**Σχεδόν Πλήρους Αναλυσιμότητας**" (NCD), η οποία είναι συνυφασμένη με τη δομή της πλειοψηφίας των ιεραρχικών συστημάτων.

## 1.2 Συνεισφορά/Οργάνωση της Διπλωματικής

Για την παρακολούθηση της παρούσας διπλωματικής εργασίας, ο αναγνώστης θεωρείται εξοικειωμένος με τις κλασσικές έννοιες γραμμικής άλγεβρας [76], θεωρίας πιθανοτήτων και αλυσίδες Markov [65].

Στο Κεφάλαιο 2 περιγράφουμε τι είναι Συνεργατική Διήθηση, τις κατηγορίες στις οποίες χωρίζεται καθώς και τα πλεονεκτήματα/μειονεκτήματα τους. Έπειτα κάνουμε μια σύντομη αναφορά στους τρόπους με τους οποίους γίνεται η επιλογή των κορυφαίων αντικειμένων στη λίστα κατάταξης που προτείνονται στο χρήστη και προχωράμε στην παρουσίαση των διαφόρων προκλήσεων που έχει να αντιμετωπίσει. Τέλος, γίνεται εκτενής αναφορά στις graph-based μεθόδους που αποτελούν τον πυρήνα της προσέγγισής μας.

Στο Κεφάλαιο 3 παρουσιάζουμε αναλυτικά τον **HIR** - το αλγοριθμικό πλαίσιο Συνεργατικής Διήθησης, που προτείνουμε[55, 56]. Αρχικά, αναφερόμαστε στην έννοια της Near Complete Decomposability (NCD) καθώς και του *NCDawareRank*[53, 54], που θεμελιώνουν διαισθητικά και τεχνικά τη μέθοδό μας. Έπειτα, αναλύουμε την κεντρική ιδέα της προσέγγισής μας και προχωράμε στη μαθηματική ανάλυση του HIR και των συστατικών του. Τέλος, παραθέτουμε τον αλγόριθμο υπολογισμού των συστάσεων και αναλύουμε την απόδοσή του, τόσο από υπολογιστικής άποψης όσο και από πλευράς αποθήκευσης.

Στο κεφάλαιο 4 εφαρμόζουμε το HIR πλαίσιο στο *movie recommendation* πρόβλημα και παραθέτουμε μια σειρά από πειράματα προκειμένου να αξιολογήσουμε την απόδοση της μεθόδου μας. Τα πειράματα αυτά πραγματοποιούνται πάνω στο MovieLens dataset. Η απόδοση σύστασης του HIR συγκρίνεται με την απόδοση άλλων αλγορίθμων κατάταξης. Τα πειράματά μας έδειξαν πως ο HIR υπερτερεί των άλλων state-of-the-art τεχνικών στις δυο ευρέως χρησιμοποιούμενες μετρικές απόδοσης αλγορίθμων συστάσεων, τις micro-DOA και macro-DOA. Επιπλέον, ελέγχουμε την απόδοση της μεθόδου μας όσον αφορά τα προβλήματα που προκαλούνται εξαιτίας της αραιότητας του χώρου αντικειμένων, διεξάγοντας έναν αριθμό από πειράματα τα οποία προσομοιώνουν το φαινόμενο αυτό. Τα πειράματα, έδειξαν πως ο HIR δεν παρουσιάζει μεγάλη ευαισθησία, ακόμα και αν η αραιότητα μεγαλώνει. Η ίδια διαπίστωση γίνεται και στην περίπτωση που έχουμε τοπική αραιότητα, η οποία συμβαίνει κατά την εισαγωγή νέων ταινιών στο σύστημα.

Τέλος, στο κεφάλαιο 5, παραθέτουμε τα συμπεράσματά μας και περιγράφουμε συνοπτικά τις κατευθύνσεις για μελλοντική έρευνα.

# Κεφάλαιο 2

# Συστήματα Συστάσεων Συνεργατικής Διήθησης

Το παρόν κεφάλαιο ακολουθεί τη διάρθρωση του "*A Survey of Collaborative Filtering Techniques*" [78]. Πιο αναλυτικά αναφέρουμε τι είναι η Συνεργατική Διήθηση, σε ποιες κατηγορίες χωρίζεται, τι πλεονεκτήματα έχουν αλλά και τι περιορισμοί εμφανίζονται που χρήζουν αντιμετώπισης. Τέλος, γίνεται εκτενής αναφορά στις graph-based μεθόδους που αποτελούν τον πυρήνα της προσέγγισής μας.

## 2.1 Συνεργατική Διήθηση

Ο όρος "Συνεργατική Διήθηση" (CF) επινοήθηκε από τους προγραμματιστές που υλοποίησαν ένα από τα πρώτα συστήματα συστάσεων, το *tapestry* [25] το οποίο χρησιμοποιήθηκε για να προτείνει έγγραφα, που προέρχονται από μια ομάδα συζήτησης, σε μια συλλογή από χρήστες. Είναι γεγονός πως αποτελεί μέχρι στιγμής την πιο επιτυχημένη τεχνολογία συστημάτων συστάσεων και χρησιμοποιείται σε πολλούς από τους πιο επιτυχημένους διαδικτυακούς τόπους ηλεκτρονικών αγορών.

Η βασική παραδοχή του CF είναι ότι αν οι χρήστες $u_i$ και $u_j$ βαθμολογήσουν $n$ αντικείμενα με παρόμοιο τρόπο ή έχουν παρόμοιες συμπεριφορές στην αλληλεπίδρασή τους με το σύστημα στο παρελθόν, τείνουν να συμφωνήσουν ξανά στο μέλλον [26]. Οι χρήστες αυτοί ονομάζονται *γείτονες* και η ομοιότητα των προτιμήσεών τους υπολογίζεται με βάση την ομοιότητα των βαθμολογιών που έδωσαν για τα αντικείμενα. Η εύρεση των γειτόνων βασίζεται στη χρήση στατιστικών τεχνικών.

Οι CF τεχνικές χρησιμοποιούν μια βάση δεδομένων που περιλαμβάνει τις προτιμήσεις των χρηστών για τα αντικείμενα. Σε ένα τυπικό CF σενάριο, υπάρχει μια λίστα από $n$ χρήστες $\{u_1, u_2, \ldots, u_n\}$ και μια λίστα από $m$ αντικείμενα $\{v_1, v_2, \ldots, v_m\}$ και κάθε χρήστης, $u_i$, έχει μια λίστα αντικειμένων, $\mathcal{I}_{u_v}$, την οποία είτε έχει βαθμολογήσει άμεσα είτε οι προ-τιμήσεις του για τα αντικείμενα συμπεραίνονται έμμεσα από την όλη συμπεριφορά του στο σύστημα. Με τον όρο "βαθμολογίες" εννοούμε είτε τις *άμεσες* είτε τις *έμμεσες* ενδείξεις:

- **Άμεσες ενδείξεις**: αποτελούν οι περιπτώσεις στις οποίες ζητείται από το χρήστη να βαθμολογήσει ένα αντικείμενο, να ταξινομήσει μια συλλογή αντικειμένων σύμφωνα με την προτίμησή του, ή να δημιουργήσει μία λίστα αντικειμένων που του αρέσουν.

- **Έμμεσες ενδείξεις**: αποτελούν η παρατήρηση των αντικειμένων από την πλοήγηση του χρήστη τα οποία βλέπει σε μια συγκεκριμένη σελίδα αντικειμένου, η δημιουργία ιστορικού με τα αντικείμενα που ο εκάστοτε χρήστης αγόρασε, καθώς και η ανάλυση του κοινωνικού δικτύου του χρήστη ώστε να εξεταστούν παρόμοιες προτιμήσεις.

Στον Πίνακα 2.1 που ακολουθεί, παρουσιάζουμε τις άμεσες βαθμολογίες που έχουν δώσει κάποιοι χρήστες για συγκεκριμένα αντικείμενα.

|         | Ροζ πάνθηρας | Εκδικητές | Μπάτμαν | Τιτανικός | Μονομάχος |
|---------|:---:|:---:|:---:|:---:|:---:|
| Μαρία   | 2 | 3 |   | 5 | 2 |
| Γιάννης | 1 | 5 | 3 | 2 |   |
| **Αλέξης** | 5 | ? | 3 | 3 | 1 |
| Άννα    | 3 | 2 | 4 |   | 2 |
| Γιώργος |   | 2 | 4 | 3 |   |

Πίνακας 2.1: Παράδειγμα ενός user-item μητρώου. Ο *Αλέξης* αποτελεί τον ενεργό χρήστη για τον οποίο το σύστημα πρέπει να κάνει συστάσεις. Σε κάποιες ταινίες δεν υπάρχουν διαθέσιμες βαθμολογίες από κάποιους χρήστες. Αυτό σημαίνει πως οι χρήστες αυτοί δεν έχουν δει ακόμα τις συγκεκριμένες ταινίες.

Η διαδικασία που ακολουθεί η συνεργατική διήθηση για την παραγωγή προβλέψεων ή συστάσεων για κάθε ενεργό χρήστη, φαίνεται στο Σχήμα 2.1.

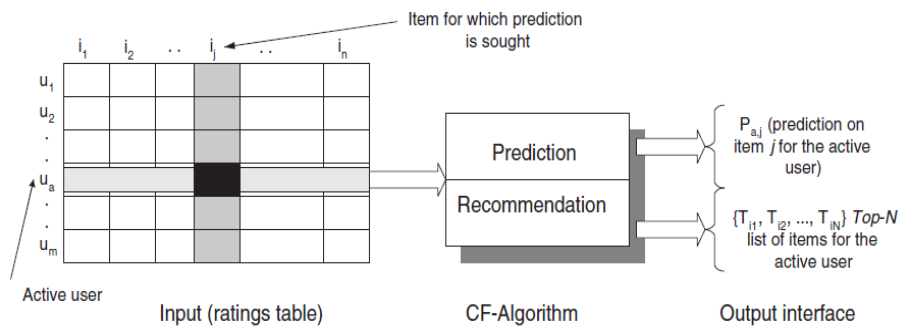

Σχήμα 2.1: Η διαδικασία της Συνεργατικής Διήθησης [70].

Παρατηρούμε πως η έξοδος του συστήματος συστάσεων λαμβάνει τη μορφή είτε μιας *πρόβλεψης* είτε μιας *σύστασης*:

- **Πρόβλεψη**: εκφράζεται σαν μια αριθμητική τιμή, η οποία αντιπροσωπεύει την εκτιμούμενη, κατά το σύστημα συστάσεων, γνώμη του ενεργού χρήστη για ένα συγκεκριμένο αντικείμενο. Η εκτιμούμενη αυτή τιμή ακολουθεί την ίδια βαθμολογική κλίμακα (πχ. από 1 μέχρι 5) με τις τιμές βαθμολογιών που έχει ήδη δώσει ο συγκεκριμένος χρήστης [70].

- **Σύσταση**: εκφράζεται υπό τη μορφή μιας λίστας $N$ αντικειμένων τα οποία θα του αρέσουν περισσότερο. Η προτεινόμενη λίστα συνήθως αποτελείται από αντικείμενα για τα οποία ο ενεργός χρήστης δεν έχει εκδηλωθεί ακόμα. Η συγκεκριμένη μορφή εξόδου των CF αλγορίθμων αναφέρεται και ως *"Top-N συστάσεις"*.

Η πρώτη γενιά των CF συστημάτων, χρησιμοποιούν τις βαθμολογίες του χρήστη για να υπολογίσουν την ομοιότητα[1] ή το βάρος μεταξύ των χρηστών ή των αντικειμένων και στη συνέχεια κάνουν προβλέψεις ή συστάσεις με βάση τις υπολογισμένες τιμές ομοιότητας. Η πρόβλεψη, λοιπόν, των βαθμολογιών για νέα αντικείμενα γίνεται με δυο τρόπους:

- **User-based**: Τα user-based συστήματα υπολογίζουν το ενδιαφέρον ενός χρήστη $u$ για ένα αντικείμενο $v$ χρησιμοποιώντας τις βαθμολογίες που έχουν δώσει οι γείτονές του. Οι γείτονες του χρήστη $u_i$ είναι τυπικά οι χρήστες $u_j$, των οποίων οι βαθμολογίες που έχουν δώσει στα κοινά αντικείμενα με τον $u_i$, είναι πολύ συσχετισμένες με εκείνες που έχει δώσει ο $u_i$. Παράδειγμα τέτοιων συστημάτων αποτελεί το GroupLens [63].

- **Item-based**: Τα item-based συστήματα προβλέπουν τη βαθμολογία ενός χρήστη $u$ για ένα αντικείμενο $v$ λαμβάνοντας υπόψη τους τις βαθμολογίες που έχει δώσει ο χρήστης $u$ σε άλλα παρόμοια αντικείμενα του $v$. Σε τέτοιες προσεγγίσεις, δυο αντικείμενα είναι όμοια αν αρκετοί χρήστες του συστήματος έχουν βαθμολογήσει αυτά τα αντικείμενα με παρόμοιο τρόπο. Παράδειγμα τέτοιων συστημάτων αποτελούν το amazon.com[2] (Σχήμα 2.2) και το Neflix[3].

Οι παραπάνω κατηγορίες αποτελούν περιπτώσεις των *memory-based CF* μεθόδων που έχουν χρησιμοποιηθεί κυρίως σε εμπορικά συστήματα, λόγω του ότι είναι αποτελεσματικοί και παράλληλα πολύ εύκολοι στην εφαρμογή τους [33, 46].

Στην επόμενη παράγραφο αναφερόμαστε πιο αναλυτικά στα memory-based συστήματα συστάσεων.

---

[1] Ο υπολογισμός της ομοιότητας γίνεται με μηχανισμούς όπως ο συντελεστής συσχέτισης Pearson ή η cosine-based ομοιότητα.

[2] www.amazon.com

[3] www.netflix.com

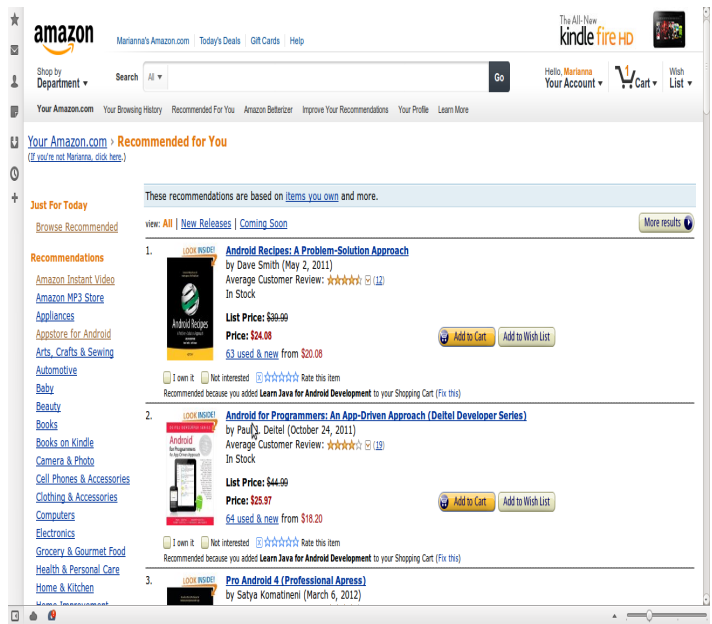

Σχήμα 2.2: Το Amazon προτείνει αντικείμενα στους χρήστες εφαρμόζοντας CF τεχνικές.

### 2.1.1 Memory-based συστήματα συστάσεων

Οι memory-based CF αλγόριθμοι χρησιμοποιούν ολόκληρη ή ένα δείγμα της user-item βάσης δεδομένων για την παραγωγή μιας πρόβλεψης. Σκοπός τους είναι να βρουν τους γείτονες ενός νέου χρήστη - εφαρμόζοντας στατιστικές τεχνικές - και να παράξουν μια πρόβλεψη όσον αφορά τις προτιμήσεις του. Η διαδικασία που ακολουθούν είναι η εξής: υπολογίζουν την ομοιότητα ή το βάρος μεταξύ δυο χρηστών (ή δυο αντικειμένων) και παράγουν μια πρόβλεψη παίρνοντας το σταθμισμένο μέσο όρο όλων των βαθμολογιών που έχει δώσει ένας συγκεκριμένος χρήστης (ή που έχει πάρει ένα συγκεκριμένο αντικείμενο) [70]. Στην περίπτωση που εξάγουν μια λίστα top-N συστάσεων, βρίσκουν τους $k$ πιο όμοιους χρήστες ή αντικείμενα (οι πιο κοντινοί γείτονες) μετά τον υπολογισμό των ομοιοτήτων, και συγκεντρώνουν τους γείτονες προκειμένου να πάρουν τα top-N πιο συχνά αντικείμενα (τα οποία αποτελούν εν τέλει την σύσταση).

#### 2.1.1.1 Υπολογισμός Ομοιότητας

Ο υπολογισμός της ομοιότητας μεταξύ αντικειμένων ή χρηστών είναι ένα πολύ σημαντικό βήμα στους memory-based CF αλγόριθμους. Για έναν *user-based CF* αλγόριθμο, πρώτα υπολογίζουμε την ομοιότητα των χρηστών $u_i$ και $u_j$ οι οποίοι έχουν βαθμολογήσει - και οι δυο - τα ίδια αντικείμενα. Αντίθετα για τους *item-based CF* αλγόριθμους, πρώτα επικεντρωνόμαστε στους χρήστες οι οποίοι έχουν βαθμολογήσει τα αντικείμενα αυτά και μετά καθορίζουμε την ομοιότητα των co-rated [1] αντικειμένων [70].

Υπάρχουν πολλές διαφορετικές μέθοδοι για τον υπολογισμό της ομοιότητας ή του βάρους μεταξύ των χρηστών ή των αντικειμένων, κάποιες από τις οποίες αναλύονται παρακάτω.

**Correlation-based ομοιότητα** Στα item-based συστήματα για να υπολογίσουμε την ομοιότητα μεταξύ δυο αντικειμένων $v_i$ και $v_j$ χρησιμοποιούμε διάφορες γνωστές μετρικές συμπεριλαμβανομένου του *συντελεστή συσχέτισης Pearson*:

$$w_{ij} \triangleq \frac{\sum_{u_k \in \mathcal{U}}(r_{ki} - \overline{r}_{v_i})(r_{kj} - \overline{r}_{v_j})}{\sqrt{\sum_{u_k \in \mathcal{U}}(r_{ki} - \overline{r}_{v_i})^2}\sqrt{\sum_{u_k \in \mathcal{U}}(r_{kj} - \overline{r}_{v_j})^2}}, \quad (2.1)$$

όπου το $r_{ki}$ είναι η βαθμολογία του χρήστη $u_k$ για το αντικείμενο $v_i$ και το $\overline{r}_{v_i}$ είναι ο μέσος όρος των βαθμολογιών που έχει πάρει το αντικείμενο $v_i$ από τους χρήστες $u_k \in \mathcal{U}$ που έχουν βαθμολογήσει και το αντικείμενο $v_j$.

Άλλες γνωστές μετρικές αυτής της κατηγορίας περιλαμβάνουν:

- τον *constrained συντελεστή συσχέτισης Pearson*, που αποτελεί μια παραλλαγή του συντελεστή Pearson η οποία χρησιμοποιεί ένα κεντρικό σημείο αντί για τη μέση βαθμολογία,

- το *συντελεστή συσχέτισης της κατάταξης του Spearman*, παρόμοια με τον συντελεστή Pearson, με τη διαφορά ότι αντί για βαθμολογίες έχουμε κατατάξεις και

- τον *Kendall's $\tau$ συντελεστή συσχέτισης* παρόμοιο με τον συντελεστή συσχέτισης της κατάταξης του Spearman, με τη διαφορά ότι αντί να χρησιμοποιηθούν οι ίδιες κατατάξεις, χρησιμοποιούνται μόνο οι σχετικές κατατάξεις για τον υπολογισμό της συσχέτισης [26, 31].

**Cosine-based ομοιότητα** Η vector cosine-based ομοιότητα, αρχικά χρησιμοποιήθηκε για τον υπολογισμό της ομοιότητας ανάμεσα σε αρχεία. Η ομοιότητα μεταξύ δυο αρχείων μπορεί να μετρηθεί αν αντιμετωπίσουμε κάθε αρχείο σαν ένα διάνυσμα των συχνοτήτων των λέξεων και υπολογίσουμε το συνημίτονο της γωνίας που σχηματίζεται από τα διανύσματα συχνότητας [67]. Ο φορμαλισμός αυτός μπορεί να υιοθετηθεί και στην περίπτωση της συνεργατική διήθησης, όπου χρήστες και αντικείμενα αντικαθιστούν τα αρχεία, και οι βαθμολογίες τις συχνότητες των λέξεων.

Τυπικά, αν $\mathbf{R}$ είναι το $n \times m$ user-item μητρώο, τότε η ομοιότητα μεταξύ δυο αντικειμένων, $v_i$ και $v_j$, ορίζεται σαν το συνημίτονο των $m$-διάστατων διανυσμάτων τα οποία αντιστοιχούν στην $i$-οστή και στην $j$-οστή στήλη του μητρώου **R**. Ένα παράδειγμα υπολογισμού για την περίπτωση της item-based ομοιότητας, φαίνεται στο Σχήμα 2.3.

---

[1] τα κοινά αντικείμενα που έχουν βαθμολογήσει δυο χρήστες $u_i$ και $u_j$.

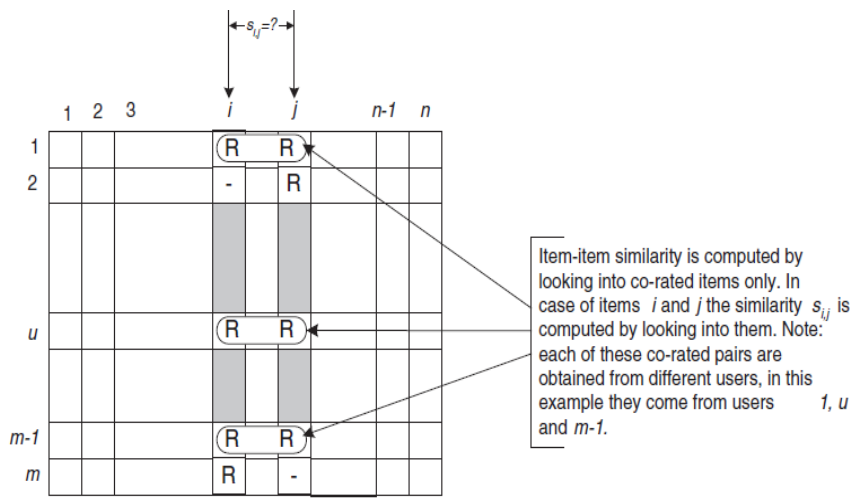

Σχήμα 2.3: Απομόνωση των co-rated αντικειμένων και ο υπολογισμός της ομοιότητας [70].

Η vector cosine ομοιότητα μεταξύ των αντικειμένων $v_i$ και $v_j$ δίνεται από τον τύπο

$$w_{ij} \triangleq \cos(\vec{v_i}, \vec{v_j}) \triangleq \frac{\vec{v_i} \bullet \vec{v_j}}{\|\vec{v_i}\| * \|\vec{v_j}\|} \tag{2.2}$$

όπου το "•" δηλώνει το εσωτερικό γινόμενο των δυο διανυσμάτων. Οπότε για την περίπτωση που έχουμε $m$ αντικείμενα, χρειάζεται να υπολογίσουμε ένα $m \times m$ μητρώο ομοιότητας [69].

**Adjusted cosine ομοιότητα**  Σε πραγματικές καταστάσεις, όπου διαφορετικοί χρήστες χρησιμοποιούν διαφορετικές κλίμακες βαθμολογίας, η vector cosine ομοιότητα δεν ενδείκνυται. Σε τέτοιες περιπτώσεις, χρησιμοποιείται η *adjusted cosine* ομοιότητα στην οποία αφαιρείται ο αντίστοιχος μέσος όρος των βαθμολογιών του χρήστη από κάθε corated ζευγάρι. Στην πραγματικότητα η συσχέτιση Pearson εκτελεί την cosine ομοιότητα με κάποιο είδος κανονικοποίησης των βαθμολογιών του χρήστη σύμφωνα με τη δική του συμπεριφορά βαθμολόγησης. Η adjusted cosine ομοιότητα δίνεται από τον τύπο:

$$w_{ij} \triangleq \frac{\sum_{u_k \in \mathcal{U}}(r_{ki} - \overline{r}_{u_k})(r_{kj} - \overline{r}_{u_k})}{\sqrt{\sum_{u_k \in \mathcal{U}}(r_{ki} - \overline{r}_{u_k})^2}\sqrt{\sum_{u_k \in \mathcal{U}}(r_{kj} - \overline{r}_{u_k})^2}} \tag{2.3}$$

όπου το $r_{ki}$ είναι η βαθμολογία που δίνει ο χρήστης $u_k$ στο αντικείμενο $v_i$, και το $\overline{r}_{u_k}$ είναι ο μέσος όρος των βαθμολογιών του χρήστη $u_k$ .

### 2.1.1.2 Top-N Συστάσεις

Η επιλογή των top-N αντικειμένων που προτείνονται στο χρήστη γίνεται με 2 τρόπους: κάνοντας *user-based* και *item-based top-N συστάσεις*. Οι τεχνικές των top-N συστάσεων αναλύουν το user-item μητρώο προκειμένου να ανακαλύψουν σχέσεις μεταξύ διαφορετικών χρηστών ή αντικειμένων και στη συνέχεια χρησιμοποιούν αυτές τις σχέσεις για να υπολογίσουν τις συστάσεις.

**User-based Top-N αλγόριθμοι συστάσεων**  Οι user-based top-N αλγόριθμοι συστάσεων αναγνωρίζουν πρώτα τους $k$ πιο όμοιους - με τον ενεργό - χρήστες χρησιμοποιώντας το συντελεστή Pearson ή το vector-space μοντέλο [7, 69]. Στα μοντέλα αυτά κάθε χρήστης συμπεριφέρεται σαν ένα διάνυσμα στο $m$-διάστατο χώρο αντικειμένων και οι ομοιότητες ανάμεσα στον ενεργό και στους άλλους χρήστες υπολογίζονται μεταξύ των διανυσμάτων, όπως αναφέραμε προηγουμένως. Από τη στιγμή που οι $k$ παρόμοιοι χρήστες έχουν ανακαλυφθεί, οι αντίστοιχες γραμμές τους στο user-item μητρώο $\mathbf{R}$ συγκεντρώνονται για να αναγνωρίσουν ένα σύνολο αντικειμένων, $\mathcal{S}$, το οποίο αγοράστηκε από παρόμοιους χρήστες. Έχοντας πλέον το σύνολο $\mathcal{S}$, οι user-based CF τεχνικές προτείνουν τα top-N πιο συχνά αντικείμενα του συνόλου $\mathcal{S}$, για τα οποία δεν έχει εκφράσει ακόμα γνώμη ο χρήστης. Παρά τη μεγάλη τους απήχηση, οι user-based top-N αλγόριθμοι συστάσεων υπόκεινται σε περιορισμούς που σχετίζονται με την επεκτασιμότητα και την real-time απόδοση [37].

**Item-based Top-N αλγόριθμοι συστάσεων**  Για την αντιμετώπιση του προβλήματος της επεκτασιμότητας που εμφανίζεται στους user-based top-N αλγόριθμους, αναπτύχθηκαν οι item-based top-N αλγόριθμοι συστάσεων. Οι αλγόριθμοι αυτοί, υπολογίζουν πρώτα, για κάθε αντικείμενο, τα $k$ πιο όμοιά του. Στη συνέχεια, προσδιορίζουν το σύνολο, $\mathcal{S}$ - το οποίο αποτελείται από τα υποψήφια αντικείμενα που θα προταθούν στο χρήστη - παίρνοντας τα $k$ πιο όμοια αντικείμενα και αφαιρώντας από αυτά το σύνολο $\mathcal{L}_i$, το οποίο περιέχει τα αντικείμενα για τα οποία ο χρήστης έχει εκφράσει άποψη. Έπειτα υπολογίζονται τις ομοιότητες μεταξύ των αντικειμένων που υπάρχουν στο σύνολο $\mathcal{S}$ και των αντικειμένων που υπάρχουν στο σύνολο $\mathcal{L}_i$. Το σύνολο $\mathcal{S}$ που προκύπτει, το οποίο είναι ταξινομημένο με φθίνουσα σειρά όσον αφορά τις ομοιότητες, αποτελεί την item-based top-N λίστα που θα προταθεί στον χρήστη [37].

Ένα πρόβλημα της μεθόδου αυτής είναι, πως όταν η από κοινού κατανομή ενός συνόλου αντικειμένων είναι διαφορετική από τις κατανομές των μεμονωμένων αντικειμένων στο σύνολο, τότε είναι πιθανόν η παραπάνω διαδικασία να οδηγήσει σε μη βέλτιστες συστάσεις [78]. Για να λυθεί το πρόβλημα αυτό, οι Deshpande και Karypis [17] ανέπτυξαν υψηλότερης τάξης item-based top-N αλγόριθμους συστάσεων, οι οποίοι χρησιμοποιούν όλους τους συνδυασμούς των αντικειμένων μέχρι ένα συγκεκριμένο μέγεθος, κατά τον καθορισμό του συνόλου των αντικειμένων που προτείνονται στο χρήστη.

### 2.1.1.3 Πλεονεκτήματα - Μειονεκτήματα

Τα memory-based συστήματα είναι πολύ εύκολα στη δημιουργία και πολύ απλά στην εφαρμογή τους. Στην πιο απλή τους μορφή, ο μόνος παράγοντας που χρειάζεται ρύθμιση είναι ο αριθμός των γειτόνων (αντικείμενα ή χρήστες) που θα χρησιμοποιηθούν για την εξαγωγή της πρόβλεψης. Ένα άλλο πλεονέκτημα είναι η σταθερότητα των συστημάτων αυτών, καθώς επηρεάζονται ελάχιστα από την συνεχή προσθήκη νέων χρηστών, αντικειμένων και βαθμολογιών, γεγονός που είναι πολύ συνηθισμένο στις μεγάλες εμπορικές εφαρμογές που τα χρησιμοποιούν. Για παράδειγμα, ένα memory-based σύστημα μπορεί να κάνει συστάσεις σε νέους χρήστες στηριζόμενο στις ομοιότητες των αντικειμένων που έχουν ήδη υπολογιστεί. Επιπλέον, εμφανίζουν άλλα δυο σημαντικά προτερήματα που σχετίζεται με την αποδοτικότητά τους: α) δε χρειάζεται να εκπαιδεύουν το σύστημα - διαδικασία ακριβή και χρονοβόρα β) η αποθήκευση των κοντινότερων γειτόνων απαιτεί πολύ μικρή μνήμη - πράγμα πολύ θετικό για τις εφαρμογές που διαχειρίζονται εκατομμύρια αντικείμενα και χρήστες. Τέλος, ένα από τα σημαντικότερα πλεονεκτήματα των memory-based συστημάτων αποτελεί το γεγονός ότι δεν απαιτούν γνώση του περιεχομένου των αντικειμένων· η γνώση των βαθμολογιών που έχουν δώσει οι χρήστες σε αυτά αρκεί.

Από την άλλη μεριά, η εξάρτηση των συστημάτων αυτών από τις βαθμολογίες των χρηστών μπορεί να οδηγήσει σε μη αποτελεσματικές προβλέψεις, αν οι χρήστες δεν έχουν αλληλεπιδράσει με αυτά. Επιπλέον, όταν τα δεδομένα των βαθμολογιών είναι αραιά - γεγονός πολύ συνηθισμένο στα συστήματα συστάσεων - η απόδοση του συστήματος μειώνεται.

Προκειμένου να ξεπεραστούν τα μειονεκτήματα των memory-based CF συστημάτων, έχουν αναπτυχθεί οι model-based προσεγγίσεις, οι οποίες αναλύονται παρακάτω.

## 2.1.2 Model-based συστήματα συστάσεων

Τα *model-based συστήματα* χρησιμοποιούν τις βαθμολογίες των χρηστών που υπάρχουν αποθηκευμένες στο σύστημα για να μάθουν ένα μοντέλο προβλέψεων. Η γενική ιδέα είναι να μοντελοποιηθούν οι user-item αλληλεπιδράσεις με παράγοντες που εκπροσωπούν τα *λανθάνοντα* χαρακτηριστικά των χρηστών και των αντικειμένων στο σύστημα, όπως η κατηγορία προτιμήσεων του χρήστη ή η κλάση στην οποία ανήκει το αντικείμενο.

Το μοντέλο αυτό *εκπαιδεύεται* χρησιμοποιώντας τα διαθέσιμα δεδομένα και στη συνέχεια εφαρμόζεται για να προβλέψει τις βαθμολογίες των χρηστών σε καινούρια αντικείμενα. Στη βιβλιογραφία υπάρχουν διάφορες model-based προσεγγίσεις για την σύσταση αντικειμένων, όπως είναι η *Bayesian Clustering* [7], η *Latent Semantic Analysis(LSI)* [32], η *Maximum Entropy* [86], οι *Boltzmann Machines* [66], οι *Support Vector Machines (SVM)* [29] και η *Singular Value Decomposition(SVD)* [4, 42, 61, 79, 80].

Στα πλεονεκτήματα των αλγορίθμων που ακολουθούν τη προσέγγιση αυτή κατατάσσεται η δυνατότητά τους να χειρίζονται αραιά δεδομένα χωρίς να μειώνεται η αποδοτικότητα του συστήματος. Επιπλέον, με την εκπαίδευση του συστήματος, μπορεί να βελτιώνεται συνεχώς η απόδοσή του. Τέλος, άλλο ένα πλεονέκτημα είναι ότι μπορούν να δίνουν μια διαισθητική εξήγηση των συστάσεων που κάνουν, και έτσι οι χρήστες εμπιστεύονται καλύτερα το σύστημα και αποδέχονται τις προτάσεις που τους γίνονται.

Από την άλλη, το βασικότερο μειονέκτημά τους είναι ότι απαιτούν ανά τακτά χρονικά διαστήματα φάσεις εκπαίδευσης για να μάθουν το μοντέλο πρόβλεψης, και η διαδικασία αυτή κοστίζει ακριβά. Επίσης, επειδή στα συστήματα αυτά χρησιμοποιούνται συχνά μέθοδοι μείωσης διαστάσεων, χρήσιμη πληροφορία μπορεί να χαθεί.

Σε μια προσπάθεια εκμετάλλευσης των πλεονεκτημάτων των δυο παραπάνω μεθόδων έχουν προταθεί στη βιβλιογραφία *υβριδικές CF τεχνικές*, οι οποίες συνδυάζουν τις CF με άλλες τεχνικές συστάσεων, συνηθέστερα με content-based προσεγγίσεις [62]. Εντούτοις, παρά τα πλεονεκτήματά τους παρουσιάζουν και κάποια μειονεκτήματα συμπεριλαμβανομένων, της αυξημένης τους πολυπλοκότητας και της δαπάνης που απαιτείται για την εφαρμογή τους.

Μία σύντομη επισκόπηση των τεχνικών του CF παρουσιάζεται στον Πίνακα 2.2.

| Κατηγορίες | Αντιπροσωπευτικές Τεχνικές | Κύρια πλεονεκτήματα | Κύρια μειονεκτήματα |
|---|---|---|---|
| **Memory-based CF** | •small Neighbor-based CF (item-based/user-based CF) αλγόριθμοι με συσχέτιση Pearson/vector cosine <br> •Item-based/user-based top-N συστάσεις | •εύκολη εφαρμογή <br> •προσθήκη νέων δεδομένων εύκολα <br> •δεν χρειάζεται εξέταση του περιεχομένου των αντικειμένων <br> •κλιμακώνεται καλά με co-rated αντικείμενα | •εξάρτηση από τις βαθμολογίες των χρηστών <br> •μειωμένη απόδοση σε αραιά δεδομένα <br> •όχι συστάσεις για νέους χρήστες και αντικείμενα <br> •περιορισμένη κλιμακωσιμότητα για μεγάλα datasets |
| **Model-based CF** | •Bayesian belief nets CF <br> •Clustering CF <br> •MDP-based CF <br> •Latent Semantic CF <br> •sparse factor analysis <br> •CF χρησιμοποιώντας τεχνικές μείωσης διαστάσεων, πχ. SVD,PCA | •καλύτερη αντιμετώπιση του sparsity, της κλιμακωσιμότητας και άλλων προβλημάτων <br> •βελτίωση απόδοσης πρόβλεψης <br> •δίνει μια διαισθητική αιτιολογία για συστάσεις | •ακριβό χτίσιμο μοντέλου <br> •trade-off μεταξύ απόδοσης και πρόβλεψης κλιμακωσιμότητας <br> •απώλεια χρήσιμης πληροφορίας για τεχνικές μείωσης διαστάσεων |
| **Hybrid Recommenders** | •content-based CF recommenders, πχ. Fab <br> •content-boosted CF <br> •hybrid CF συνδυάζοντας memory-based και model-based CF αλγόριθμους, πχ. Personal Diagnosis | •ξεπερνά τους περιορισμούς των CF και content-based ή άλλων recommenders <br> •βελτίωση απόδοσης πρόβλεψης <br> •ξεπερνά περιορισμούς του CF όπως το sparsity και το και το gray sheep | •αυξημένη πολυπλοκότητα δαπάνη για την εφαρμογή <br> •ανάγκη για εξωτερική πληροφορία που δεν είναι διαθέσιμη |

Πίνακας 2.2: Πλεονεκτήματα και Μειονεκτήματα τεχνικών Συνεργατικής Διήθησης.

## 2.2 Προκλήσεις του CF

Εξαιτίας του γεγονότος πως τα CF συστήματα συστάσεων είναι ευρέως διαδεδομένα, πολλές προκλήσεις με τις οποίες έρχονται αντιμέτωπα, πρέπει να εξεταστούν. Οι CF αλγόριθμοι απαιτείται να έχουν την ικανότητα να αντεπεξέλθουν με πολύ αραιά δεδομένα, να μπορούν να προσαρμοστούν στον αυξανόμενο αριθμό των χρηστών και των αντικειμένων, να κάνουν ικανοποιητικές συστάσεις σε μικρό χρονικό διάστημα και να αντιμετωπίζουν άλλα προβλήματα όπως είναι η συνωνυμία, οι επιθέσεις κακόβουλων χρηστών, ο θόρυβος των δεδομένων και η ιδιωτικότητα των προσωπικών δεδομένων.

Πριν προχωρήσουμε στην ανάλυσή τους, είναι σημαντικό να επισημάνουμε πώς, συνήθως, ένα RS το οποίο πετυχαίνει γρήγορες και ακριβείς συστάσεις, μπορεί να επιφέρει όχι μόνο την ικανοποίηση του πελάτη αλλά και θετικά αποτελέσματα στις εταιρίες που το εφαρμόζουν. Συνεπώς, για τα CF συστήματα, η παραγωγή υψηλής ποιότητας προβλέψεων και συστάσεων, εξαρτάται από το πόσο καλά θα αντιμετωπιστούν οι παρακάτω προκλήσεις.

**Αραιά δεδομένα** Όπως αναφέραμε και σε προηγούμενη ενότητα, τα συστήματα συστάσεων συνεργατικής διήθησης βασίζονται στις ομοιότητες μεταξύ χρηστών όπως αυτές προκύπτουν από τη σύγκριση των βαθμολογιών τους για τα ίδια αντικείμενα. Οι χρήστες όμως είναι πολύ πιθανό να μην έχουν βαθμολογήσει κοινά αντικείμενα ή να έχουν βαθμολογήσει μόνο λίγα. Το user-item μητρώο συνήθως προκύπτει υπερβολικά αραιό κάτι που όπως είναι αναμενόμενο θέτει σε "δοκιμασία" την πλειοψηφία των CF συστημάτων.

Η πρόκληση των *αραιών δεδομένων* εμφανίζεται σε αρκετές περιπτώσεις:

- **cold start**: πρόβλημα συμβαίνει όταν ένας νέος χρήστης ή ένα νέο αντικείμενο έχει μόλις εισέλθει στο σύστημα. Στην περίπτωση αυτή, είναι δύσκολο να βρεθούν παρόμοιοι χρήστες ή αντικείμενα, επειδή δεν υπάρχουν επαρκείς πληροφορίες. Το cold start πρόβλημα, επίσης αναφέρεται και ως *πρόβλημα νέου χρήστη ή πρόβλημα νέου αντικειμένου* [1, 83].Από τη μια μεριά, νέα αντικείμενα δεν μπορούν να προταθούν μέχρις ότου βαθμολογηθούν πρώτα από κάποιους χρήστες, και από την άλλη είναι αδύνατον να δοθούν καλές συστάσεις στους χρήστες λόγω της έλλειψης των βαθμολογιών τους και του ιστορικού των αγορών τους.

- **κάλυψη** (coverage): μπορεί να καθοριστεί σαν το ποσοστό των αντικειμένων για τα οποία ο αλγόριθμος θα μπορούσε να παρέχει συστάσεις. Το πρόβλημα της *περιορισμένης κάλυψης* (reduced coverage) συμβαίνει όταν ο αριθμός των βαθμολογιών που έχουν δώσει οι χρήστες είναι πολύ μικρός σε σχέση με τον αριθμό των αντικειμένων που υπάρχουν στο σύστημα.

- **neighbor transitivity** : αναφέρεται σαν ένα πρόβλημα το οποίο συμβαίνει σε αραιές

βάσεις δεδομένων, όπου δυο χρήστες με παρόμοια γούστα είναι πιθανόν να μην αναγνωριστούν ως παρόμοιοι, αν δεν έχουν βαθμολογήσει και οι δυο τα ίδια αντικείμενα.

Για την εξάλειψη του προβλήματος των αραιών δεδομένων, έχουν προταθεί στη βιβλιογραφία μια σειρά από προσεγγίσεις. Τεχνικές μείωσης διαστάσεων (dimensionality reduction), όπως είναι ο SVD [6], προβάλλουν τους χρήστες και τα αντικείμενα σε ένα μητρώο μειωμένων διαστάσεων στο χώρο, το οποίο συμπεριλαμβάνει τα πιο βασικά χαρακτηριστικά τους. Με αυτόν τον τρόπο, σε αυτόν τον χώρο πυκνών χαρακτηριστικών, μπορούν να βρεθούν σχέσεις ακόμα και ανάμεσα σε χρήστες που δεν έχουν βαθμολογήσει τα ίδια αντικείμενα. Η μείωση διαστάσεων γίνεται είτε στο user-item μητρώο, είτε στο μητρώο αποθήκευσης των ομοιοτήτων. Μια μέθοδος η οποία βασίζεται στον SVD και χρησιμοποιείται ευρέως σε περιπτώσεις ανάκτησης πληροφορίας είναι η *Latent Semantic Indexing* (LSI) [16, 45], όπου η ομοιότητα μεταξύ των χρηστών καθορίζεται από την αναπαράστασή τους. Οι Goldberg et al. [26] ανέπτυξαν τον *Eigentaste*, έναν CF αλγόριθμο που εφαρμόζει την PCA τεχνική, η οποία περιγράφτηκε πρώτα από τον Pearson το 1901, για μείωση διαστάσεων. Εντούτοις, σε τέτοιες περιπτώσεις, όταν απορρίπτονται συγκεκριμένοι χρήστες ή αντικείμενα, χάνονται και χρήσιμες πληροφορίες για συστάσεις που σχετίζονται με αυτούς, πράγμα το οποίο συνεπάγεται υποβάθμιση της ποιότητας των συστάσεων [46, 69]. Υβριδικοί CF αλγόριθμοι, όπως ο content-boosted CF αλγόριθμος [49], εμφανίστηκαν στο προσκήνιο προκειμένου να αντιμετωπιστεί το πρόβλημα της αραιότητας των δεδομένων. Στις περιπτώσεις αυτές, για την παραγωγή προβλέψεων για νέους χρήστες ή νέα αντικείμενα, χρησιμοποιούνται εξωτερικές πληροφορίες περιεχομένου.

Οι Kim και Li [40] πρότειναν ένα πιθανοτικό μοντέλο για την αντιμετώπιση του cold start προβλήματος, στο οποίο τα αντικείμενα ταξινομούνται σε ομάδες και οι προβλέψεις που γίνονται για τους χρήστες λαμβάνουν υπόψη την Gaussian κατανομή των βαθμολογιών των χρηστών. Επιπλέον model-based CF αλγόριθμοι, έχουν προταθεί επίσης στη βιβλιογραφία, όπως είναι ο TAN-ELR [30, 77], και αντιμετωπίζουν το πρόβλημα της αραιότητας παρέχοντας πιο ακριβείς προβλέψεις για τα αραιά δεδομένα. Τέλος, ένας άλλος τρόπος χειρισμού των προβλημάτων της περιορισμένης κάλυψης και των αραιών δεδομένων αποτελούν οι graph-based μέθοδοι, τους οποίους αναλύουμε σε επόμενη ενότητα, λόγω της συνάφειας με τον δικό μας αλγόριθμο.

**Επεκτασιμότητα** Όταν ο αριθμός των χρηστών και των αντικειμένων μεγαλώνει υπερβολικά, οι παραδοσιακοί CF αλγόριθμοι υποφέρουν από σοβαρά προβλήματα επεκτασιμότητας, με υπολογιστικούς πόρους πέρα από τα πρακτικά και αποδεχτά επίπεδα. Επίσης, πολλά συστήματα είναι ανάγκη να αντιδράσουν άμεσα σε online απαιτήσεις και να κάνουν συστάσεις για όλους τους χρήστες, άσχετα με το ιστορικό των αγορών τους και τις βαθμολογίες που έχουν δώσει στα αντικείμενα.

## 2. ΣΥΝΕΡΓΑΤΙΚΗ ΔΙΗΘΗΣΗ

Τεχνικές μείωσης διαστάσεων, όπως η SVD, μπορούν σε ένα βαθμό να αντιμετωπίσουν το πρόβλημα της επεκτασιμότητας και να παράγουν γρήγορα καλής ποιότητας συστάσεις. Ωστόσο, η χρήση του συνεπάγεται ακριβά βήματα προκειμένου να γίνει η παραγοντοποίηση του μητρώου. Ένας αυξητικός SVD CF αλγόριθμος [68] προϋπολογίζει την SVD χρησιμοποιώντας ήδη υπάρχοντες χρήστες. Όταν ένα νέο σύνολο βαθμολογιών προστίθεται στη βάση δεδομένων, ο αλγόριθμος χρησιμοποιεί την *folding-in* τεχνική προβολής [5, 16] για να δημιουργήσει ένα αυξητικό σύστημα χωρίς να χρειαστεί να υπολογίσει το χαμηλών διαστάσεων μοντέλο από το μηδέν. Συνεπώς, το σύστημα συστάσεων επεκτείνεται εύκολα.

Επιπλέον, model-based CF αλγόριθμοι, όπως είναι οι αλγόριθμοι ομαδοποίησης CF, προσπαθούν να λύσουν το πρόβλημα ψάχνοντας για χρήστες που θα τους κάνουν συστάσεις, μέσα σε μικρότερες παρόμοιες ομάδες χρηστών και όχι σε ολόκληρη τη βάση δεδομένων [11, 59, 71, 82]. Στην περίπτωση αυτή πρέπει να σημειωθεί πως υπάρχουν tradeoffs μεταξύ της επεκτασιμότητας και της απόδοσης της πρόβλεψης.

**Συνωνυμία** Η *συνωνυμία* αναφέρεται στην τάση που έχουν κάποια ίδια ή πολύ παρόμοια αντικείμενα, να έχουν διαφορετικά ονόματα ή καταχωρήσεις στη βάση δεδομένων. Τα περισσότερα συστήματα συστάσεων, δεν είναι ικανά να ανακαλύψουν αυτή τη κρυμμένη συσχέτιση και ως εκ τούτου αντιμετωπίζουν αυτά τα αντικείμενα με διαφορετικό τρόπο. Για παράδειγμα, τα φαινομενικά διαφορετικά αντικείμενα "children movie" και "children film" είναι στην πραγματικότητα το ίδιο αντικείμενο, αλλά τα memory-based CF συστήματα δεν μπορούν να βρουν κάποιο ταίριασμα ανάμεσά τους ώστε να υπολογίσουν την μεταξύ τους ομοιότητα. Είναι γεγονός πως το πρόβλημα της συνωνυμίας μειώνει την απόδοση των CF συστημάτων.

Προηγούμενες προσπάθειες για την επίλυση του προβλήματος συνωνυμίας, βασίστηκαν στην αυτόματη επέκταση όρου ή ακόμα και στην κατασκευή ενός θησαυρού. Το μειονέκτημα των πλήρως αυτόματων μεθόδων, ωστόσο, είναι πως κάποιοι προστιθέμενοι όροι μπορεί να έχουν διαφορετικές σημασίες από αυτές για τις οποίες προορίζονται. Συνεπώς, αυτό οδηγεί σε γρήγορη υποβάθμιση της απόδοσης των συστάσεων [35].

Οι SVD τεχνικές, συγκεκριμένα η LSI μέθοδος, είναι ικανές να αντιμετωπίσουν προβλήματα που οφείλονται στη συνωνυμία. Η SVD παίρνει ένα μεγάλο term-document μητρώο δεδομένων συσχέτισης και δημιουργεί ένα σημασιολογικό χώρο στον οποίο όροι και έγγραφα που είναι στενά συσχετισμένα, τοποθετούνται κοντά το ένα στο άλλο. Η SVD επιτρέπει την διάταξη του χώρου για να αντικατοπτρίζει τα σημαντικά πρότυπα συσχέτισης στα δεδομένα, και να αγνοεί τα μικρότερα και λιγότερο σημαντικά. Η απόδοση του LSI στην αντιμετώπιση του προβλήματος συνωνυμίας είναι εντυπωσιακή σε υψηλότερα επίπεδα ανάκλησης (recall), όπου η ακρίβεια (precision) είναι συνήθως πολύ χαμηλή [16].

Τέλος, η LSI μέθοδος, δίνει μόνο μερική λύση στο πρόβλημα της *πολυσημίας* (polysemy), η οποία αναφέρεται στο γεγονός ότι οι πιο πολλές λέξεις έχουν περισσότερες από μια διακριτή έννοια [16].

**Gray Sheep - Black Sheep**  Ο όρος *gray sheep* αναφέρεται σε μια μικρή ή μεσσαία κοινωνία χρηστών, όπου υπάρχουν άτομα τα οποία δεν επωφελούνται καθόλου από το απλό συνεργατικό φιλτράρισμα, εξαιτίας του γεγονότος ότι οι απόψεις τους δεν έχουν κάποια συνέπεια με τις απόψεις οποιασδήποτε ομάδας ανθρώπων. Τα άτομα αυτά λαμβάνουν σπάνια, αν όχι ποτέ, ακριβείς συστάσεις ακόμα και μετά από την αρχική φάση εκκίνησης για το χρήστη και για το σύστημα [12].

Από την άλλη μεριά, ο όρος *black sheep* αναφέρεται σε μια ομάδα ανθρώπων όπου είναι δύσκολο να τους γίνουν συστάσεις, εξαιτίας του γεγονότος ότι δεν υπάρχουν καθόλου ή υπάρχουν πολύ λίγα άτομα, τα οποία σχετίζονται μαζί τους λόγω των ιδιαίτερων προτιμήσεων που έχουν (βλέπε Σχήμα 2.4).

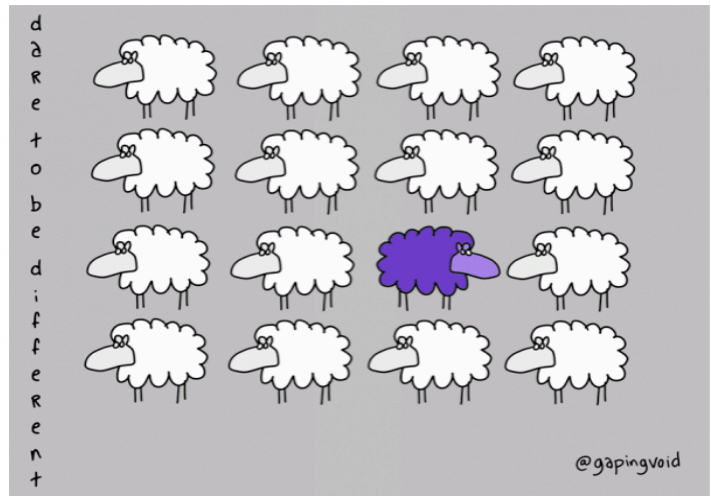

Σχήμα 2.4: Το "μαύρο πρόβατο".

Οι Claypool et al. [12] παρέχουν μια υβριδική προσέγγιση που βασίζει την πρόβλεψη στο σταθμισμένο μέσο όρο της content-based και της CF πρόβλεψης. Σε αυτή την προσέγγιση, τα βάρη των content-based και CF προβλέψεων καθορίζονται για κάθε χρήστη.

**Επιθέσεις κακόβουλων χρηστών**  Σε περιπτώσεις όπου ο καθένας μπορεί να παρέχει συστάσεις, οι χρήστες είναι πιθανόν να δώσουν πάρα πολλές θετικές συστάσεις για το δικό τους υλικό και αρνητικές συστάσεις για τους ανταγωνιστές τους. Συνεπώς, οι επιθέσεις αυτές μπορούν να συμβάλλουν στην ενίσχυση ενός αντικειμένου ή στο αντίθετο. Είναι επιθυμητό λοιπόν, τα CF συστήματα να εισάγουν μέτρα ασφαλείας τα οποία αποθαρρύνουν τέτοιου είδους φαινόμενα [64].

Πρόσφατα, για ένα CF σύστημα, έχουν αναγνωριστεί κάποια *μοντέλα κακόβουλων χρηστών* και η αποτελεσματικότητά τους έχει μελετηθεί. Ο Lam και Riedl [44] βρήκαν πως ένας item-based CF αλγόριθμος επηρεάζεται πολύ λιγότερο από τις επιθέσεις σε σχέση με τον user-based CF αλγόριθμο. Οι Mobasher et al. [51] έχουν προχωρήσει στην εξέταση μοντέλων επιθέσεων για τα item-based CF συστήματα, και επιπλέον πιστεύουν πως εναλλακτικά CF συστήματα, όπως είναι τα υβριδικά CF και τα model-based CF συστήματα, είναι δυνατόν να παρέχουν μερικές λύσεις στο *bias injection πρόβλημα*. Επιπλέον, οι O'Mahony et al. [58] συνέβαλαν στη λύση του προβλήματος των επιθέσεων από κακόβουλους χρήστες αναλύοντας την ευρωστία - δηλαδή την προσαρμοστικότητα ενός RS σε πιθανές κακόβουλες διαταράξεις που μπορεί να συμβούν στο user-item μητρώο βαθμολογιών.

**Άλλες προκλήσεις** Εξαιτίας του γεγονότος πως κάποιοι άνθρωποι δεν θέλουν οι συνήθειές τους και οι προτιμήσεις τους να είναι ευρέως γνωστές, τα CF συστήματα γεννούν επίσης την ανησυχία για την ιδιωτικότητα των προσωπικών δεδομένων. Οι Miller et al. [50] και Canny [10] βρήκαν τρόπους για να διαφυλάξουν την ιδιωτικότητα των χρηστών στην περίπτωση των CF συστάσεων.

Ο αυξανόμενος θόρυβος είναι μια άλλη πρόκληση, καθώς ο πληθυσμός των χρηστών γίνεται πιο ποικίλος. Οι *από κοινού MMMF* [15] και *instance selection* τεχνικές [84] βρέθηκαν χρήσιμες για την αντιμετώπιση του θορύβου όσον αφορά τα CF ζητήματα. Παρά το γεγονός ότι η θεωρία Dempster-Shafer (DS) [9, 72] και οι imputation τεχνικές [39] έχουν εφαρμοστεί με επιτυχία για να δέχονται τα ατελή και θορυβώδη δεδομένα, για την αναπαράσταση της γνώσης αλλά και για ζητήματα ομαδοποίησης, εντούτοις μπορούν να χρησιμοποιηθούν και για την επίλυση του προβλήματος αυτού.

Η έννοια "explainability" είναι μια άλλη σημαντική πτυχή των συστημάτων συστάσεων. Η παραγωγή μιας διαισθητικής εξήγησης, όπως "θα αρέσει σε κάποιον αυτό το βιβλίο, επειδή του άρεσαν αυτά τα βιβλία" θα είναι ελκυστική και ωφέλιμη για τους αναγνώστες, ανεξάρτητα από την ακρίβεια των εξηγήσεων [42].

## 2.3 Graph-based τεχνικές

Έναν άλλον τρόπο χειρισμού των προβλημάτων της περιορισμένης κάλυψης και των αραιών δεδομένων, όπως αναφέρθηκε προηγουμένως, αποτελούν οι *graph-based* μέθοδοι. Στις προσεγγίσεις που ανήκουν σε αυτή την κατηγορία, τα δεδομένα αναπαρίστανται με τη μορφή γράφου όπου οι κόμβοι είναι χρήστες, αντικείμενα ή και τα δυο μαζί και οι ακμές αντιστοιχούν στις σχέσεις ή στις ομοιότητες μεταξύ των χρηστών και των αντικειμένων. Για παράδειγμα, στο Σχήμα 2.5, τα δεδομένα μοντελοποιούνται με ένα διμερή γράφο, όπου τα 2 σύνολα των κόμβων αναπαριστούν χρήστες και αντικείμενα, και μια ακμή του γράφου συνδέει ένα χρήστη $u$ με ένα αντικείμενο $v$, αν ο χρήστης $u$ έχει βαθμολογήσει το αντικείμενο $v$. Στην ακμή αυτή είναι δυνατόν να δοθεί ένα βάρος, όπως είναι η βαθμολογία που

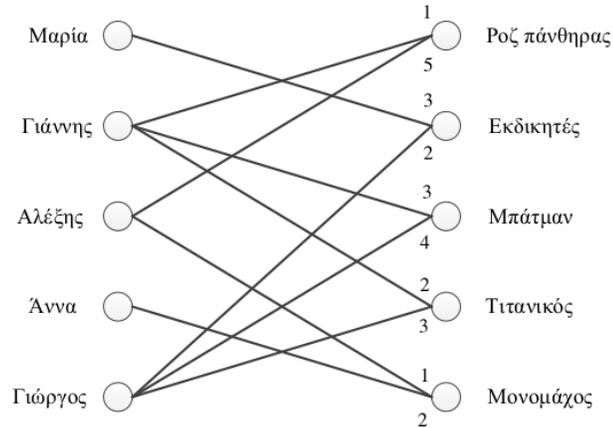

Σχήμα 2.5: Διμερής γράφος. Τα βάρη στις ακμές αντιπροσωπεύουν βαθμολογίες χρηστών για τα αντικείμενα.

έχει δώσει ο χρήστης. Σε ένα άλλο μοντέλο γράφου, όπου οι κόμβοι αναπαριστούν είτε χρήστες, είτε αντικείμενα, η ακμή που συνδέει δυο κόμβους δηλώνει το βαθμό συσχέτισής τους.

Οι graph-based προσεγγίσεις, επιτρέπουν σε κόμβους που δεν συνδέονται απευθείας με μια ακμή να επηρεάζουν ο ένας τον άλλον, καθώς θεωρούν ότι η πληροφορία μεταφέρεται στις συνδεόμενες ακμές. Όσο μεγαλύτερο είναι το βάρος μιας ακμής, τόσο περισσότερη πληροφορία επιτρέπεται να περάσει μέσω αυτής. Επίσης, η επιρροή ενός κόμβου σε έναν άλλον, μικραίνει όσο πιο απομακρυσμένοι είναι οι κόμβοι μεταξύ τους. Αυτές οι δυο ιδιότητες, γνωστές ως *propagation* και *attenuation* [27, 34], παρατηρούνται συχνά σε graph-based μέτρα ομοιότητας.

Η εκμετάλλευση των μεταβατικών σχέσεων των δεδομένων σε ένα γράφο μειώνει τα προβλήματα της περιορισμένης κάλυψης και των αραιών δεδομένων, αφού πλέον μπορούν να εκτιμηθούν σχέσεις ανάμεσα σε χρήστες ή αντικείμενα που δεν συνδέονται άμεσα μεταξύ τους. Αυτές οι μεταβατικές σχέσεις χρησιμοποιούνται για την σύσταση αντικειμένων με δυο τρόπους. Στη μια περίπτωση, η γειτνίαση ενός χρήστη με ένα αντικείμενο στο γράφο χρησιμοποιείται απευθείας για να εκτιμηθεί η βαθμολογία που θα έδινε ο χρήστης $u$ στο αντικείμενο $v$ [20, 27, 34]. Πρακτικά, αυτό σημαίνει ότι στον $u$ προτείνονται τα αντικείμενα εκείνα που του είναι πιο κοντά στον γράφο. Στην δεύτερη περίπτωση, ο αλγόριθμος λαμβάνει υπόψη του τη γειτνίαση δυο κόμβων-χρηστών ή κόμβων-αντικειμένων ως μέτρο ομοιότητας, και στη συνέχεια χρησιμοποιεί αυτή την ομοιότητα σαν βάρη στη neighbor-based μέθοδο συστάσεων [20, 47]. Είναι γεγονός πως υπάρχουν διάφοροι τρόποι υπολογισμού της ομοιότητας στους γράφους, κάποιοι από τους οποίους αναφέρονται παρακάτω.

### 2.3.1 Ομοιότητα με βάση το μονοπάτι

Στην ομοιότητα με βάση το μονοπάτι (path-based) η απόσταση μεταξύ δυο κόμβων, εκτιμάται ως συνάρτηση του αριθμού των μονοπατιών τα οποία συνδέουν τους δυο κόμβους και του μήκους αυτών των μονοπατιών.

**Αριθμός μονοπατιών** Ο αριθμός των μονοπατιών μεταξύ ενός χρήστη και ενός αντικειμένου σε ένα διμερή γράφο μπορεί επίσης να χρησιμοποιηθεί για να υπολογίσει την συμβατότητά τους [34]. Έστω $\mathbf{R}$, για άλλη μια φορά, ένα $|U| \times |V|$ μητρώο που περιέχει τις βαθμολογίες που έχουν δώσει οι χρήστες για τα αντικείμενα. Ισχύει ότι, το $r_{uv}$ ισούται με 1 αν ο χρήστης $u$ έχει βαθμολογήσει το αντικείμενο $v$, αλλιώς ισούται με 0. Το μητρώο γειτνίασης $\mathbf{A}$ του διμερούς γράφου μπορεί, λοιπόν, να οριστεί από το μητρώο $\mathbf{R}$ σαν

$$\mathbf{A} = \begin{pmatrix} \mathbf{O} & \mathbf{R}^T \\ \mathbf{R} & \mathbf{O} \end{pmatrix}$$

Στην προσέγγιση αυτή λοιπόν, η συσχέτιση ανάμεσα σε ένα χρήστη $u$ και κάποιο αντικείμενο $v$ ορίζεται ως το άθροισμα των βαρών όλων των διαφορετικών μονοπατιών που συνδέουν τους κόμβους $u$ και $v$ (επιτρέποντας στους κόμβους να εμφανίζονται περισσότερες από μια φορά σε κάθε μονοπάτι), των οποίων το μήκος δεν είναι μεγαλύτερο από μια δεδομένη μέγιστη τιμή $K$. Πρέπει να σημειωθεί ότι, εφόσον ο γράφος είναι διμερής, το $K$ πρέπει να είναι περιττός αριθμός. Προκειμένου να μετριάσει τη συμβολή των μεγαλύτερων μονοπατιών, το βάρος που δίνεται σε ένα μονοπάτι μήκους $k$ ορίζεται ως $\alpha^k$, όπου $\alpha \in [0,1]$. Από τη στιγμή που ο αριθμός των μονοπατιών, μήκους $k$, μεταξύ των ζευγαριών των κόμβων, δίνεται από το $\mathbf{A}^k$, το user-item μητρώο συσχέτισης $\mathbf{S}_k$ είναι:

$$\begin{aligned} \mathbf{S}_k &= \sum_{k=1}^{K} \alpha^k \mathbf{A}^k \\ &= (\mathbf{I} - \alpha \mathbf{A})^{-1}(\alpha \mathbf{A} - \alpha^K \mathbf{A}^K). \end{aligned}$$

Η μέθοδος αυτή, η οποία υπολογίζει αποστάσεις μεταξύ των κόμβων του γράφου, είναι γνωστή ως μέτρο Katz [38]. Πρέπει να σημειωθεί πως, το μέτρο αυτό είναι στενά συσχετισμένο με το Von Neumann Diffusion kernel [21, 41, 43]

$$\begin{aligned} \mathbf{K}_{VND} &= \sum_{k=0}^{\infty} \alpha^k \mathbf{A}^k \\ &= (\mathbf{I} - \alpha \mathbf{A})^{-1} \end{aligned}$$

και το Exponential Diffusion Kernel

$$\begin{aligned} \mathbf{K}_{ED} &= \sum_{k=0}^{\infty} \frac{1}{k!} \alpha^k \mathbf{A}^k \\ &= exp(\alpha \mathbf{A}), \end{aligned}$$

όπου $\mathbf{A}^0 = \mathbf{I}$.

Στα συστήματα συστάσεων τα οποία έχουν μεγάλο αριθμό χρηστών και αντικειμένων, ο υπολογισμός αυτών των τιμών συσχέτισης είναι πιθανόν να απαιτήσει πάρα πολλούς υπολογιστικούς πόρους. Προκειμένου να ξεπεραστούν αυτοί οι περιορισμοί, έχουν χρησιμοποιηθεί στο [34] εκτεταμένες *τεχνικές ενεργοποίησης* (activation techiques) [14]. Ουσιαστικά, τέτοιες τεχνικές δουλεύουν ενεργοποιώντας πρώτα ένα επιλεγμένο υποσύνολο κόμβων σαν κόμβους εκκίνησης, και στη συνέχεια ενεργοποιώντας επαναληπτικά τους κόμβους τους οποίους φθάνουν οι κόμβοι που είναι ήδη ενεργοί, μέχρι να συναντηθεί ένα κριτήριο σύγκλισης.

**Κοντινότερο μονοπάτι** Στην περίπτωση της συνάρτησης του μήκους των μονοπατιών, η ομοιότητα υπολογίζεται βάσει της μικρότερης απόστασής τους στον γράφο (δηλαδή όσο πιο κοντά βρίσκονται 2 χρήστες στο γράφο, τόσο πιο όμοιοι είναι) [2]. Τα δεδομένα μοντελοποιούνται σαν έναν κατευθυνόμενο γράφο του οποίου οι κόμβοι είναι χρήστες, και οι ακμές του καθορίζονται με βάση τις έννοιες *horting* και *predictability*. Horting είναι μια σχέση ανάμεσα σε δυο χρήστες που ικανοποιείται όταν οι χρήστες έχουν βαθμολογήσει όμοια αντικείμενα. Τυπικά, ένας χρήστης $u_i$ "horts" έναν άλλον χρήστη $u_j$ υπό την προϋπόθεση ότι ικανοποιείται είτε η σχέση $|\mathcal{I}_{u_i u_j}| \geqslant \alpha$ είτε η σχέση $|\mathcal{I}_{u_i u_j}|/|\mathcal{I}_{u_j}| \geqslant \beta$, όπου $\alpha$, $\beta$ είναι προκαθορισμένα κατώτατα όρια. Predictability, από την άλλη μεριά, είναι μια ισχυρότερη έννοια, που απαιτεί οι βαθμολογίες δυο χρηστών $u_i$ και $u_j$ να είναι παρόμοιες, αφού έχει θεωρηθεί πρώτα και η διαφορά στην βαθμολογική κλίμακα των δυο χρηστών.

### 2.3.2 Random walk ομοιότητα

Άλλη μέθοδος υπολογισμού της ομοιότητας είναι η *random walk* ομοιότητα, στην οποία οι μεταβατικές σχέσεις στον γράφο ορίζονται σε ένα πιθανοτικό πλαίσιο. Η ομοιότητα ανάμεσα σε δυο κόμβους σε αυτήν την προσέγγιση εκτιμάται ως η πιθανότητα να συναντήσει κανείς αυτούς τους κόμβους σε έναν τυχαίο περίπατο. Αυτό μαθηματικοποιείται με μια πρώτης τάξης αλυσίδα Markov με μητρώο πιθανοτήτων μετάβασης $\mathbf{P} \in \mathbb{R}^{n \times n}$.

Έστω $\boldsymbol{\pi}(t)$ διάνυσμα το οποίο περιέχει την *κατανομή καταστάσεων t βημάτων*. Τότε η εξέλιξη της Μαρκοβιανής αλυσίδας χαρακτηρίζεται από

$$\boldsymbol{\pi}(t+1) = \boldsymbol{\pi}(t)\mathbf{P}.$$

Επιπλέον, υπό την προϋπόθεση ότι το μητρώο $\mathbf{P}$ είναι στοχαστικό κατά γραμμές, δηλαδή $\sum_j p_{ij} = 1$ για όλα τα $i$, η διαδικασία συγκλίνει σε μια *κατανομή ισορροπίας* $\boldsymbol{\pi}(\infty)$ που αποτελεί το αριστερό ιδιοδιάνυσμα του $\mathbf{P}$ που αντιστοιχεί στην ιδιοτιμή 1. Η διαδικασία αυτή περιγράφεται συχνά με τη μορφή ενός γράφου με βάρη έχοντας έναν κόμβο για κάθε κατάσταση, και όπου η πιθανότητα μετάβασης από έναν κόμβο σε έναν γειτονικό δίνεται από το βάρος της ακμής η οποία συνδέει τους δυο αυτούς κόμβους.

**Itemrank** Μια προσέγγιση σύστασης, η οποία βασίζεται στον αλγόριθμο PageRank ο οποίος χρησιμοποιείται για την ταξινόμηση των σελίδων του διαδικτύου [8], είναι ο ItemRank [27]. Η προσέγγιση αυτή ταξινομεί τις προτιμήσεις ενός χρήστη $u$ για νέα αντικείμενα $i$, ως την πιθανότητα του $u$ να επισκεφτεί το $i$ σε ένα τυχαίο περίπατο στο γράφο όπου οι κόμβοι αντιστοιχούν στα αντικείμενα του συστήματος και οι ακμές συνδέουν αντικείμενα τα οποία έχουν βαθμολογηθεί από κοινούς χρήστες. Τα βάρη των ακμών δίνονται από το $|I|\times|I|$ μητρώο *πιθανοτήτων μετάβασης* $\mathbf{P}$ για το οποίο το $p_{ij} = |U_{ij}|/|\mathcal{U}_i|$ είναι η υπολογισμένη δεσμευμένη πιθανότητα ενός χρήστη να βαθμολογήσει ένα αντικείμενο $j$ αν έχει βαθμολογήσει ένα αντικείμενο $i$.

Όπως γίνεται στον PageRank, ο τυχαίος περιηγητής μπορεί, οποιαδήποτε στιγμή $t$, είτε να μεταπηδήσει σε έναν γειτονικό κόμβο χρησιμοποιώντας το μητρώο $\mathbf{P}$ με μια καθορισμένη πιθανότητα $a$, ή να "τηλεμεταφερθεί" σε οποιοδήποτε κόμβο με πιθανότητα $(1-a)$. Έστω ότι η $\mathbf{r}_u$ είναι η $u$-οστή γραμμή του μητρώου βαθμολογίας $\mathbf{R}$, η κατανομή πιθανότητας του χρήστη $u$ να "τηλεμεταφερθεί" σε άλλους κόμβους δίνεται από το διάνυσμα $\mathbf{d}_u = \mathbf{r}_u/\|\mathbf{r}_u\|$.

Ακολουθώντας αυτούς τους ορισμούς, το διάνυσμα στάσιμης κατανομής[1] του χρήστη $u$ στο βήμα $t+1$ μπορεί να εκφραστεί αναδρομικά ως

$$\boldsymbol{\pi}_u(t+1) = a\mathbf{P}^{\mathsf{T}}\boldsymbol{\pi}_u(t) + (1-a)\mathbf{d}_u \tag{2.4}$$

Για πρακτικούς λόγους, το $\boldsymbol{\pi}(\infty)$ συνήθως αποκτιέται με μια διαδικασία η οποία πρώτα αρχικοποιεί την κατανομή ως *ομοιόμορφη*, δηλαδή $\boldsymbol{\pi}_u(0) = \frac{1}{n}\mathbf{1}_n$, και στη συνέχεια επαναληπτικά ενημερώνει την $\boldsymbol{\pi}_u$ χρησιμοποιώντας την 2.4, μέχρι να συγκλίνει. Εφόσον η $\boldsymbol{\pi}_u(\infty)$ έχει υπολογιστεί, το σύστημα προτείνει στο χρήστη $u$ το αντικείμενο $v$ για το οποίο το σκορ κατάταξης είναι υψηλότερο.

---

[1] Το μητρώο πιθανοτήτων μετάβασης σε αυτή την περίπτωση είναι στοχαστικό κατά στήλες.

**Average first-passage/commute time**   Άλλα μέτρα απόστασης, τα οποία βασίζονται στις τυχαίες περιηγήσεις, έχουν προταθεί για τα προβλήματα συστάσεων. Ανάμεσα σε αυτά είναι το *average first-passage time* και το *average commute time*, τα οποία έχουν προταθεί από τους Fouss et al. [20, 21].

Το average first-passage time (χρόνος πρώτης προσπέλασης) $m(j \mid i)$ [57] είναι ο μέσος αριθμός των βημάτων που χρειάζονται από έναν τυχαίο περιηγητή για να φτάσει σε έναν κόμβο $j$ για πρώτη φορά, όταν ξεκινά από τον κόμβο $i \neq j$. Αν $\mathbf{P}$ είναι το $n \times n$ μητρώο πιθανοτήτων μετάβασης, τότε το $m(j \mid i)$ μπορεί να εκφραστεί αναδρομικά ως

$$m(j \mid i) = \begin{cases} \emptyset & , \text{ if } i = j \\ 1 + \sum_{k=1}^{n} p_{ik} m(j \mid k) & , \text{ αλλιώς} \end{cases} \quad (2.5)$$

Ένα πρόβλημα με το average first-passage time είναι ότι δεν είναι συμμετρικό. Ένα σχετικό μέτρο το οποίο δεν αντιμετωπίζει αυτό το πρόβλημα είναι το average commute time $n(i, j) = m(j \mid i) + m(i \mid j)$ [24], το οποίο αντιστοιχεί στο μέσο αριθμό των βημάτων που απαιτούνται από τον τυχαίο περιηγητή ο οποίος ξεκινά από τον κόμβο $i \neq j$ να φτάσει στον κόμβο $j$ για πρώτη φορά και μετά να γυρίσει πίσω στον $i$. Το μέτρο αυτό έχει αρκετές ενδιαφέρουσες ιδιότητες. Συγκεκριμένα, πρόκειται για ένα αληθινό μέτρο απόστασης στον Ευκλείδειο χώρο [24] το οποίο σχετίζεται με τη γνωστή ιδιότητα των αντιστάσεων στα ηλεκτρικά κυκλώματα και το ψευδό-αντίστροφο Laplacian μητρώο που παράγεται από το γράφο [20].

Στο [20], το average commute-time χρησιμοποιείται για να υπολογίσει την απόσταση μεταξύ των κόμβων ενός διμερούς γράφου, οι οποίοι αναπαριστούν τις αλληλεπιδράσεις των χρηστών και των αντικειμένων σε ένα σύστημα συστάσεων. Για κάθε χρήστη $u$ υπάρχει μια κατευθυνόμενη ακμή από τον $u$ σε κάθε αντικείμενο $i \in \mathcal{I}_u$, και το βάρος αυτής της ακμής είναι απλά $1/|\mathcal{I}_u|$. Παρομοίως, υπάρχει μια κατευθυνόμενη ακμή από κάθε αντικείμενο $i$ σε κάθε χρήστη $u \in \mathcal{U}_i$, με βάρος $1/|\mathcal{U}_i|$. Το average commute time μπορεί να χρησιμοποιηθεί με δυο διαφορετικούς τρόπους: Πρώτον, με το να προτείνει στον $u$ το αντικείμενο $i$ για το οποίο το $n(u, i)$ είναι το μικρότερο, και δεύτερον με το να βρει τους χρήστες οι οποίοι είναι οι πιο κοντινοί στον χρήστη $u$, σύμφωνα με την commute time distance, και μετά να προτείνει στον $u$ τα αντικείμενα τα οποία άρεσαν περισσότερο σε αυτούς τους χρήστες.

Ανακεφαλαιώνοντας, παρουσιάσαμε την ιδέα της συνεργατικής διήθησης την οποία εφαρμόζουν τα συστήματα συστάσεων προκειμένου να παράξουν προβλέψεις ή συστάσεις για τους χρήστες που αλληλεπιδρούν μαζί τους. Επιπλέον, αναφέραμε τις προκλήσεις που παρουσιάζονται στα συστήματα αυτά, δίνοντας παράλληλα λύσεις για την αντιμετώπιση τους και στο τέλος αφιερώσαμε μια ολόκληρη ενότητα στις graph-based μεθόδους, λόγω του ότι σε αυτές βασίζεται και η δική μας προσέγγιση. Στο επόμενο κεφάλαιο παρουσιάζουμε την αναλυτική περιγραφή της μεθόδου μας.

# 2. ΣΥΝΕΡΓΑΤΙΚΗ ΔΙΗΘΗΣΗ

# Μέρος II

# Σύστημα Συστάσεων HIR

# Κεφάλαιο 3

# HIR Framework

Στο παρόν κεφάλαιο προτείνουμε ένα νέο σύστημα συστάσεων - τον HIR - και παρουσιάζουμε τη μοντελοποίησή του, χτίζοντας πάνω στη διαίσθηση πίσω από τον αλγόριθμο NCDawareRank [53]. Συγκεκριμένα, υπολογίζουμε ένα εξατομικευμένο διάνυσμα κατάταξης, εκμεταλλευόμενοι τόσο τις άμεσες συσχετίσεις μεταξύ των αντικειμένων όσο και την αναλυσιμότητα (decomposability) του χώρου κατάστασής τους.

## 3.1 Near Complete Decomposability: Η διαίσθηση πίσω από τον HIR

Τα μεγάλα συστήματα που εμφανίζονται στη φύση, στην πλειοψηφία τους, δεν είναι τόσο σύνθετα όσο το μέγεθός τους υπονοεί. Αντιθέτως, ο χώρος κατάστασής τους είναι σχεδόν άδειος και τα σχετικά μητρώα που τα περιγράφουν έχουν την τάση να είναι αραιά, και "δομημένα". Σύμφωνα με τον Νομπελίστα Herbert A. Simon [73], αυτή η εγγενής αραιότητα είναι συνυφασμένη με την εξελικτική βιωσιμότητα των συστημάτων αυτών και την δομική τους οργάνωση. Υποστήριξε πως η πλειοψηφία των ιεραρχικά δομημένων σύνθετων συστημάτων, μοιράζονται την ιδιότητα της *σχεδόν πλήρους αναλυσιμότητας (nearly completely decomposable, NCD)*: οι καταστάσεις τους οργανώνονται σε ιεραρχικά επίπεδα από μπλοκ, υπο-μπλοκ, υπο-υπο-μπλοκ κ.ο.κ, με τέτοιο τρόπο ώστε οι αλληλεπιδράσεις ανάμεσα στα στοιχεία που ανήκουν στο ίδιο μπλοκ να είναι πολύ πιο ισχυρές από τις αλληλεπιδράσεις μεταξύ στοιχείων που ανήκουν σε διαφορετικά μπλοκ.

Το γεγονός ότι ένα σύνθετο σύστημα μπορεί να κατέχει την ιδιότητα της NCD σηματοδοτεί τα εξής:

- δείχνει το δρόμο προς μια κατάλληλη προσέγγιση μοντελοποίησης και μια μαθηματική ανάλυση, η οποία τονίζει τα ενδημικά χαρακτηριστικά του συστήματος,

- μπορεί να βοηθήσει στην ανακούφιση των προβλημάτων που προέρχονται από την

- αραιότητα του υποκείμενου χώρου κατάστασής,

- δίνει μια βαθύτερη γνώση της συμπεριφοράς του και συνεπώς,

- παρέχει ένα εννοιολογικό πλαίσιο για την ανάπτυξη των αλγορίθμων και των μεθόδων τα οποία εκμεταλλεύονται αυτή τη γνώση από μια ποιοτική και υπολογιστική οπτική γωνία.

Η πρώτη ανάλυση NCD συστημάτων αποδίδεται στους Simon και Ando [74], που αναφέρθηκαν στην ομαδοποίηση καταστάσεων σε γραμμικά μοντέλα οικονομικών συστημάτων. Ωστόσο, η ευελιξία της ιδέας του Simon έχει επιτρέψει την εφαρμογή της NCD με εμφανή επιτυχία σε πολλά σύνθετα συστήματα στον τομέα της βιολογίας, των κοινωνικών επιστημών, του management, της εξέλιξης κ.τ.λ. Η εισαγωγή της NCD στην επιστήμη των υπολογιστών οφείλεται στον Courtois [13], ο οποίος πέτυχε να θεμελιώσει μαθηματικά την ιδέα, και να την εφαρμόσει με χαρακτηριστική επιτυχία στη θεωρία ουρών αναμονής και στην ανάλυση απόδοσης υπολογιστικών συστημάτων. Τα τελευταία χρόνια, η NCD έχει χρησιμοποιηθεί στο πρόβλημα κατάταξης ιστοσελίδων, πρώτα από υπολογιστικής πλευράς [19, 36, 81] και έπειτα από ποιοτικής άποψης [53, 54].

**NCDawareRank** Πρόκειται για έναν αλγόριθμο κατάταξης ιστοσελίδων, ο οποίος προτάθηκε πρόσθετα στη βιβλιογραφία [53, 54] προκειμένου να γενικεύσει αλλά και να βελτιώσει το γνωστό αλγόριθμο PageRank [60]. Εκμεταλλευόμενος την NCD ιδιότητα που έχει ο Ιστός και διατηρώντας τα - από μαθηματικής πλευράς ελκυστικά - χαρακτηριστικά του PageRank, παράγει καλύτερα αποτελέσματα τόσο από ποιοτική όσο και από υπολογιστική άποψη. Επιπλέον, καταφέρνει να λύσει με μεγάλη επιτυχία πολλά από τα γνωστά προβλήματα του διαδικτύου συμπεριλαμβανομένων, του Web Spamming και της μεροληψίας των νεοεισερχόμενων σελίδων.

## 3.2 Ορισμοί και Διαίσθηση

Το σύστημά μας αποτελείται από χρήστες και αντικείμενα. Το σύνολο των *χρηστών* συμβολίζεται με $\mathcal{U} = \{u_1, \ldots, u_n\}$ και το σύνολο των *αντικειμένων* με $\mathcal{V} = \{v_1, \ldots, v_m\}$. Θεωρούμε μια διαμέριση $\{\mathcal{L}, \mathcal{T}\}$ των βαθμολογιών σε ένα σύνολο εκπαίδευσης (training set) και σε ένα σύνολο ελέγχου (test set). Για κάθε χρήστη $u_i$, ορίζουμε το $\mathcal{L}_i$ ως το σύνολο των ταινιών που υπάρχουν στο training set και τις οποίες ο χρήστης έχει βαθμολογήσει, και το $\mathcal{T}_i$ ως το σύνολο των ταινιών που υπάρχουν στο test set και τις οποίες ο χρήστης έχει, επίσης, βαθμολογήσει. Πιο τυπικά:

$$\mathcal{L}_i \triangleq \{v_k \mid t_{ik} \in \mathcal{L}\} \quad \text{και} \quad \mathcal{T}_i \triangleq \{v_l \mid t_{il} \in \mathcal{T}\}$$

Επιπλέον, κάθε χρήστης $u_i$ σχετίζεται με ένα διάνυσμα $\boldsymbol{\omega} \triangleq [\omega_1, \omega_2, \ldots, \omega_m]$ του οποίου τα μη μηδενικά στοιχεία περιέχουν τις βαθμολογίες του χρήστη που περιλαμβάνονται στο training set $\mathcal{L}$ και το οποίο είναι κανονικοποιημένο ώστε τα στοιχεία του να αθροίζουν στη μονάδα. Το διάνυσμα αυτό στο εξής θα αναφέρεται, *διάνυσμα προτίμησης του χρήστη*.

Ορίζουμε, ακόμα, το $\mathcal{R}$ σαν ένα σύνολο τριπλετών, $t_{ij} = (u_i, v_j, r_{ij})$, όπου το $r_{ij}$ είναι ένας μη αρνητικός αριθμός που αναπαριστά τη "βαθμολογία" που δίνει ο χρήστης $u_i$ στο αντικείμενο $v_j$.

Τα αντικείμενα χαρακτηρίζονται μοναδικά από κάποιες ιδιότητες. Αν για παράδειγμα θεωρήσουμε τις "ταινίες" σαν αντικείμενα, τότε οι ταινίες χαρακτηρίζονται από την κατηγορία στην οποία ανήκουν (όπου μπορεί να είναι 1 ή περισσότερες), από τον σκηνοθέτη τους, από τους σεναριογράφους τους κ.ο.κ. Στη γενική περίπτωση των αντικειμένων, θεωρούμε μια οικογένεια συνόλων $\mathcal{D} = \{\mathcal{D}_1, \mathcal{D}_2, \ldots, \mathcal{D}_K\}$ η οποία ορίζει ένα *decomposition* του χώρου των αντικειμένων, $\mathcal{V}$, σύμφωνα με κάποια κριτήρια, όπως είναι για παράδειγμα η κατηγορία μιας ταινίας. Κάθε σύνολο $\mathcal{D}_I$ αναφέρεται σαν ένα *NCD μπλοκ*.

Επιπλέον ορίζουμε το $\mathcal{X}_v$ ως το *NCD σύνολο εγγύτητας* (NCD-proximal σύνολο) των αντικειμένων $v \in \mathcal{V}$. Το σύνολο αυτό υπολογίζεται ως η ένωση των NCD μπλοκ τα οποία περιέχουν το αντικείμενο $v$. Τυπικά, το σύνολο $\mathcal{X}_v$ δίνεται από την εξίσωση:

$$\mathcal{X}_v \triangleq \bigcup_{v \in \mathcal{D}_k} \mathcal{D}_k \tag{3.1}$$

Προκειμένου, να υποδηλώσουμε τον αριθμό των διαφορετικών μπλοκ που υπάρχουν στο $\mathcal{X}_v$, χρησιμοποιούμε το $N_v$. Με άλλα λόγια, ο αριθμός $N_v$ ενός αντικειμένου $v$ είναι ίσος με τον αριθμό των μπλοκ στα οποία ανήκει το αντικείμενο αυτό. Στο σημείο αυτό, είναι σημαντικό να τονιστεί πως το *decomposition* που ορίζεται από το $\mathcal{D}$, δεν χρειάζεται να είναι μια διαμέριση. Η μόνη απαίτηση είναι ότι $\mathcal{V} = \bigcup_{k=1}^{K} \mathcal{D}_k$, δηλαδή, ίσως υπάρχουν ακέραιοι $I, J$ τέτοιοι ώστε $\mathcal{D}_I \cap \mathcal{D}_J \neq \emptyset$. Αυτό σημαίνει πως μπορεί να υπάρχουν αντικείμενα τα οποία ανήκουν σε παραπάνω από 1 μπλοκ, όπως αναφέρθηκε προηγουμένως.

Κυρίαρχος στόχος μας είναι να δημιουργήσουμε μια λίστα κατάταξης, που θα περιέχει τις προτιμήσεις κάθε χρήστη του συστήματος. Στη διαδικασία αυτή, είναι πολύ σημαντικό το πλαίσιο που προτείνουμε να μπορέσει να παράξει τη λίστα αυτή, αντιμετωπίζοντας παράλληλα και την ευαισθησία που οφείλεται στα αραιά δεδομένα. Για το λόγο αυτό, χρησιμοποιούμε τη διαίσθηση πίσω από τον NCDawareRank αλγόριθμο [53] - ο οποίος υπολογίζει διανύσματα κατάταξης ιεραρχικά δομημένων γράφων - εφαρμόζοντας τη σχετική με αυτόν έννοια της NCD εγγύτητας.

Ακολούθως, παρουσιάζουμε τον τρόπο με τον οποίο εκμεταλλευόμαστε την πληροφορία που παρέχεται τόσο από τις *άμεσες σχέσεις* των αντικειμένων όσο και από το *decomposability* του χώρου τους.

Έχοντας ορίσει τις παραμέτρους του μοντέλου μας, τώρα είμαστε έτοιμοι να εισάγουμε ένα εξατομικευμένο στοχαστικό μητρώο $\mathbf{B}(\boldsymbol{\omega}^{(i)}) \in \mathbb{R}^{m \times m}$, όπου το $\boldsymbol{\omega}^{(i)}$ αποτελεί την κατανομή προτίμησης ενός συγκεκριμένου χρήστη πάνω σε όλα τα αντικείμενα. Το μητρώο αυτό πρέπει να αντανακλά την άμεση, καθώς επίσης και την decomposable δομή του χώρου των αντικειμένων, προκειμένου να παράγει εξατομικευμένες συστάσεις. Συνεπώς, το μητρώο $\mathbf{B}$ αποτελείται από 3 συστατικά, το *μητρώο άμεσης συσχέτισης* $\mathbf{C}$, το *μητρώο NCD εγγύτητας* $\mathbf{D}$ και το *μητρώο προτίμησης χρήστη* $\mathbf{E}$ το οποίο προέρχεται από τις άμεσες βαθμολογίες που δίνει ο χρήστης στα αντικείμενα.

Το συνολικό μητρώο $\mathbf{B}$ ορίζεται επομένως ως εξής:

$$\mathbf{B} \triangleq \alpha\mathbf{C} + \beta\mathbf{D} + (1-\alpha-\beta)\mathbf{E}$$

με τα $\alpha, \beta$ να είναι θετικοί πραγματικοί αριθμοί τέτοιοι ώστε, $\alpha + \beta < 1$.

Στην ενότητα που ακολουθεί, παραθέτουμε την ανάλυση των τριών αυτών συστατικών.

## 3.3 HIR μητρώα

Στην ενότητα αυτή αναλύονται τα 3 μητρώα τα οποία αποτελούν το στοχαστικό μητρώο $\mathbf{B}$, το οποίο παράγει εξατομικευμένες συστάσεις στους χρήστες.

**Μητρώο άμεσης συσχέτισης** Η δημιουργία του μητρώου άμεσης συσχέτισης $\mathbf{C}$ έχει ως σκοπό να ανακαλυφθούν οι άμεσες σχέσεις των αντικειμένων $u \in \mathcal{V}$. Γενικά, κάθε τέτοιο αντικείμενο θα συνδέεται με μια διακριτή κατανομή $\mathbf{c}_v = [c_1, c_2, \cdots, c_m]$ πάνω στο $\mathcal{V}$, η οποία αντανακλά τη συσχέτιση μεταξύ αυτών των αντικειμένων. Υπάρχουν διάφοροι τρόποι για να ορίσουμε αυτή την κατανομή, οι οποίοι εξαρτώνται από την φύση του προβλήματος, τις πληροφορίες που είναι διαθέσιμες κ.τ.λ. Πρέπει να σημειωθεί πως, από μαθηματικής άποψης, η μόνη απαίτηση του framework μας είναι ότι το μητρώο $\mathbf{C}$ είναι στοχαστικό κατά γραμμές και συνεπώς εδώ το κριτήριο για την επιλογή της κατανομής δεν παίζει ρόλο. Στην περίπτωσή μας, καθώς και για όλα τα πειράματα που παρουσιάζονται στην επόμενη ενότητα, χρησιμοποιούμε το σταθμισμένο μέσο όρο (α) μιας ελαφριάς τροποποίησης του μητρώου συσχέτισης που χρησιμοποιείται από τον ItemRank [27] και (β) μιας κανονικοποιημένης κατά γραμμές έκδοσης του κλασικού adjusted cosine μητρώου ομοιότητας [70]. Τα μητρώα αυτά περιγράφονται αυστηρά παρακάτω.

Αρχικά, αξίζει να αναφέρουμε πως επιθυμούμε να εξάγουμε την "ιδέα" της συσχέτισης άμεσα από τις προτιμήσεις του χρήστη σαν ένα σύνολο. Στην πραγματικότητα, ένας χρήστης τείνει στο να έχει αρκετές ομογενείς προτιμήσεις γύρω από τα αντικείμενα. Συνεπώς μπορούμε πολύ λογικά να σκεφτούμε πώς αν το αντικείμενο $v_i$ και το αντικείμενο $v_j$ έχουν την τάση να εμφανίζονται σε πολλές λίστες προτιμήσεων διαφορετικών χρηστών, τότε το $v_i$ και το $v_j$ σχετίζονται. Συνεπώς, η συσχέτιση που ψάχνουμε συνδέεται με ένα *co-occurrence* κριτήριο. Από τη στιγμή που καταφέρουμε να εκμεταλλευτούμε αυτή την πληροφορία, τότε μπορούμε αρκετά εύκολα να υπολογίσουμε τις προτιμήσεις του εξαρτημένου χρήστη.

Ορίζουμε, λοιπόν, ένα μητρώο $\mathbf{U}$, $\mathcal{V} \times \mathcal{V}$ διαστάσεων, του οποίου τα στοιχεία είναι $|\mathcal{U}_{ij}|$, όπου το $\mathcal{U}_{ij} \subseteq \mathcal{U}$ υποδηλώνει το σύνολο των χρηστών οι οποίοι έχουν βαθμολογήσει και το αντικείμενο $v_i$ και το αντικείμενο $v_j$, δηλαδή:

$$\mathcal{U}_{ij} \triangleq \begin{cases} \{u_k : (v_i \in \mathcal{L}_k) \wedge (v_j \in \mathcal{L}_k)\} & \text{αν } i \neq j \\ \emptyset & \text{αλλιώς} \end{cases}, \quad (3.2)$$

όπου το $\mathcal{L}_k$ είναι το σύνολο των βαθμολογιών που έχει δώσει ο συγκεκριμένος χρήστης για τα αντικείμενα του συστήματος. Είναι προφανές πώς το στοιχείο $\mathcal{U}_{ii} = 0$ και το μητρώο $\mathbf{U}$ είναι συμμετρικό. Προχωράμε στην κανονικοποίηση αυτού του μητρώου, ώστε να αποκτήσουμε ένα στοχαστικό κατά γραμμές μητρώο[1], $\widehat{\mathbf{U}}$. Το μητρώο $\widehat{\mathbf{U}}$ ονομάζεται *μητρώο συσχέτισης*, και κάθε είσοδός του περιέχει το δείκτη συσχέτισης μεταξύ των ζευγαριών των αντικειμένων. Το μητρώο αυτό μπορεί να αναπαρασταθεί επίσης και σαν γράφος συσχέτισης $G_\mathcal{U}$. Οι κόμβοι στο γράφο $G_\mathcal{U}$ αντιστοιχούν στα αντικείμενα που ανήκουν στο σύνολο $\mathcal{V}$ και υπάρχει μια ακμή $(v_i,v_j)$ αν και μόνο αν $\widehat{\mathbf{U}}_{ij} > 0$. Επιπλέον το βάρος το οποίο σχετίζεται με την ακμή $(v_i,v_j)$ θα είναι $\widehat{\mathbf{U}}_{ij}$. Όμως από τη στιγμή που το μητρώο έγινε στοχαστικό, παύει να είναι συμμετρικό. Δηλαδή, το βάρος της ακμής $(v_i,v_j)$ είναι διαφορετικό από το βάρος της ακμής $(v_j,v_i)$.

Ο γράφος συσχέτισης είναι ένα πολύτιμο μοντέλο γράφου, το οποίο είναι χρήσιμο για την αξιοποίηση της συσχέτισης που υπάρχει μεταξύ των αντικειμένων, και η οποία φαίνεται από τα βάρη που ανατίθενται στις ακμές του με βάση την πληροφορία που εξάγεται από τις βαθμολογίες των χρηστών. Το προκύπτον μητρώο, λοιπόν, θα οριστεί ως ακολούθως:

$$\mathbf{H} \triangleq \widehat{\mathbf{U}} + \mathbf{d}\boldsymbol{\omega}^{\mathsf{T}} \quad (3.3)$$

όπου το $\mathbf{d}$ είναι ένα διάνυσμα το οποίο υποδεικνύει τις μηδενικές γραμμές του μητρώου $\mathbf{U}$ (δηλαδή το $i^{th}$ στοιχείο του είναι 1 αν και μόνο αν $U_{ij} = 0$ για κάθε $j$) και το $\boldsymbol{\omega}^{\mathsf{T}}$ είναι το κανονικοποιημένο κατά γραμμές διάνυσμα προτίμησης του χρήστη. Είναι λογικό να υπάρχουν οι μηδενικές γραμμές, διότι σε πραγματικά συστήματα υπάρχουν αντικείμενα τα οποία δεν σχετίζονται με κάποιο άλλο αντικείμενο. Αυτό συνήθως ισχύει για τα καινούρια αντικείμενα που εισάγονται στο σύστημα. Συνεπώς, το τελικό μητρώο $\mathbf{H}$ γίνεται ένα στοχαστικό μητρώο.

Όσον αφορά το δεύτερο μητρώο άμεσης συσχέτισης που χρησιμοποιούμε στην προσέγγισή μας, για τον υπολογισμό της συσχέτισης των αντικειμένων, εφαρμόζουμε την adjusted cosine ομοιότητα. Όπως ειπώθηκε στην Ενότητα 2.1.1.1, σε πραγματικές κατα-στάσεις όπου διαφορετικοί χρήστες χρησιμοποιούν διαφορετικές κλίμακες βαθμολογίας, η adjusted cosine ομοιότητα, στην οποία αφαιρείται ο αντίστοιχος μέσος όρος των βαθμολογιών του χρήστη από κάθε co-rated ζευγάρι, δίνει πιο αξιόπιστο αποτέλεσμα σε σχέση με τις άλλες cosine-based ομοιότητες.

---

[1] στοχαστικό κατά γραμμές μητρώο ονομάζεται το μητρώο του οποίου κάθε μη μηδενική γραμμή αθροίζεται στη μονάδα.

Συνεπώς, το adjusted cosine μητρώο ομοιότητας ορίζεται ακολούθως ως:

$$\mathbf{S} \triangleq \widehat{\mathbf{G}} + \mathbf{d}'\boldsymbol{\omega}^\intercal \quad (3.4)$$

όπου το μητρώο $\widehat{\mathbf{G}}$ είναι η κανονικοποιημένη κατά γραμμές έκδοση του μητρώου $\mathbf{G}$, του οποίου το $ij^{th}$ στοιχείο ορίζεται παρακάτω:

$$G_{ij} \triangleq \frac{\sum_{u_k \in \mathcal{U}}(r_{ki} - \overline{r}_{u_k})(r_{kj} - \overline{r}_{u_k})}{\sqrt{\sum_{u_k \in \mathcal{U}}(r_{ki} - \overline{r}_{u_k})^2}\sqrt{\sum_{u_k \in \mathcal{U}}(r_{kj} - \overline{r}_{u_k})^2}} \quad (3.5)$$

Συγκεκριμένα, το $\overline{r}_{u_k}$ είναι ο μέσος όρος των βαθμολογιών του χρήστη $u_k$, και το $r_{ki}$ είναι η βαθμολογία που δίνει ο χρήστης $u_k$ στο αντικείμενο $v_i$.

Τελικά, το *συνολικό μητρώο άμεσης συσχέτισης* που προκύπτει είναι: $\mathbf{C} \triangleq \phi\mathbf{H} + (1-\phi)\mathbf{S}$, όπου $\phi$ είναι θετικός ακέραιος και μικρότερος από τη μονάδα.

**Μητρώο NCD εγγύτητας** Το μητρώο NCD εγγύτητας $\mathbf{D}$, ορίζει μια "μεταξύ επιπέδων" σύνδεση των στοιχείων του χώρου αντικειμένων. Η βασική ιδέα είναι πως η βαθμολογία ενός χρήστη, εκτός του ότι εκφράζει την άμεση γνώμη του για ένα συγκεκριμένο αντικείμενο, δίνει επιπλέον μια ένδειξη γύρω από τη γνώμη του για το NCD proximal σύνολο αυτού του αντικειμένου. Για παράδειγμα, αν αρέσει σε ένα χρήστη ένα αντικείμενο το οποίο ανήκει σε ένα ή περισσότερα μπλοκ, τότε είναι πιθανό να του αρέσουν και τα άλλα αντικείμενα που ανήκουν σε αυτά τα μπλοκ. Συνεπώς, η "εξάπλωση" των συνολικών επιδράσεων κάθε ατομικής γνώμης, μέσω των NCD decompositions, σε πολλά σχετιζόμενα αντικείμενα στο χώρο των αντικειμένων, είναι δυνατόν να συμβάλλει στον υπολογισμό της πραγματικής συσχέτισης των αντικειμένων και επίσης να βοηθήσει στην αντιμετώπιση των προβλημάτων που σχετίζονται με τα αραιά δεδομένα.

Ακολουθώντας την κατεύθυνση αυτή, σχετίζουμε κάθε γραμμή του μητρώου $\mathbf{D}$ με ένα διάνυσμα πιθανότητας $\mathbf{d}_v$, το οποίο κατανέμει ομοιόμορφα τη μάζα του ανάμεσα στα $N_v$ μπλοκ του $\mathcal{X}_v$, και στη συνέχεια ομοιόμορφα στα αντικείμενα που περιλαμβάνονται σε κάθε μπλοκ. Τυπικά, η $i$-οστή γραμμή του μητρώου αυτού, σχετίζει το αντικείμενο $v_i$ με κάθε άλλο αντικείμενο του συνόλου $\mathcal{X}_{v_i}$ όπως ορίζεται από τη σχέση (3.1). Το $ij$ στοιχείο του μητρώου $\mathbf{D}$, το οποίο σχετίζει το αντικείμενο $v_i$ με το αντικείμενο $v_j$ ορίζεται ως:

$$D_{ij} \triangleq \sum_{\mathcal{D}_k \in \mathcal{X}_{v_i}, v_j \in \mathcal{D}_k} \frac{1}{N_{v_i}|\mathcal{D}_k|} \quad (3.6)$$

Από τους ορισμούς των NCD μπλοκ και των NCD proximal συνόλων, γίνεται διαισθητικά προφανές πώς οποτεδήποτε το $K < m$, δηλαδή ο αριθμός των μπλοκ είναι μικρότερος από τον αριθμό των αντικειμένων, το μητρώο $\mathbf{D}$ είναι ένα χαμηλής τάξης (low rank) μητρώο. Εκμεταλλευόμενοι την ιδέα πίσω από την παραγοντοποίηση του interlevel proximity μητρώου $\mathbf{M}$ που χρησιμοποιείται από τον NCDawareRank [53] βλέπουμε πως μπορούμε να γράψουμε το μητρώο $\mathbf{D}$ με τρόπο ώστε να πετύχουμε αποτελεσματική αποθήκευση και υπολογισιμότητα.

Συγκεκριμένα, ορίζουμε ένα μητρώο συνάθροισης $\mathbf{A} \in \mathbb{R}^{m \times K}$, του οποίου το $ij^{th}$ στοιχείο είναι 1, αν $v_i \in \mathcal{D}_j$, αλλιώς 0. Συνεπώς η παραγοντοποίηση που μπορεί να γίνει στο μητρώο $\mathbf{D}$ είναι η ακόλουθη:

$$\mathbf{D} = \mathbf{XY} \tag{3.7}$$

όπου το $\mathbf{X}$ και το $\mathbf{Y}$ είναι οι κανονικοποιημένες κατά γραμμές εκδόσεις του $\mathbf{A}$ και του $\mathbf{A}^\intercal$ αντίστοιχα.

Για την διευκόλυνση του αναγνώστη και την καλύτερη κατανόηση των παραπάνω ορισμών, παραθέτουμε ένα απλό παράδειγμα, που φαίνεται στο Σχήμα 3.1.

$$\begin{array}{c}
\begin{array}{ccccc} \mathcal{D}_1 & \mathcal{D}_2 & \mathcal{D}_3 & \mathcal{N}_v & \mathcal{X}_v \end{array} \\
\begin{array}{c} v_1 \\ v_2 \\ v_3 \\ v_4 \\ v_5 \\ v_6 \end{array}
\left( \begin{array}{ccc|c|c}
\checkmark & - & - & 1 & \{v_1, v_2, v_4, v_6\} \\
\checkmark & \checkmark & - & 2 & \{v_1, v_2, v_4, v_5, v_6\} \\
- & - & \checkmark & 1 & \{v_3, v_6\} \\
\checkmark & \checkmark & - & 2 & \{v_1, v_3, v_4, v_5, v_6\} \\
- & \checkmark & - & 1 & \{v_2, v_4, v_5\} \\
\checkmark & - & \checkmark & 2 & \{v_1, v_2, v_3, v_4, v_6\}
\end{array} \right)
\end{array}$$

Σχήμα 3.1: Decomposition του χώρου 6 αντικειμένων σε 3 NCD μπλοκ

Στο παράδειγμα αυτό, ο χώρος των αντικειμένων, $\mathcal{V}$, αποτελείται από 6 αντικείμενα, τα οποία γίνονται αναλύονται σε 3 NCD μπλοκ. Η τελευταία στήλη του πίνακα περιέχει τα υπολογισμένα σύνολα εγγύτητας. Για παράδειγμα, το σύνολο $\mathcal{X}_{v_2}$ των αντικειμένων που είναι "κοντά" στο $v_2$ είναι $\mathcal{D}_1 \cup \mathcal{D}_2 \equiv \{v_1, v_2, v_5, v_6\}$. Αναλυτικά το παραπάνω βγαίνει ως εξής: Παρατηρούμε ότι το αντικείμενο $v_2$ ανήκει σε 2 μπλοκ, στο $\mathcal{D}_1$ και στο $\mathcal{D}_2$. Στο $\mathcal{D}_1$ μπλοκ ανήκουν όμως και τα αντικείμενα $v_1$, $v_4$ και $v_6$ και στο $\mathcal{D}_2$ ανήκουν επίσης το $v_4$ και το $v_5$ εκτός από το $v_2$. Συνεπώς, όλα τα αντικείμενα που ανήκουν σε ένα τουλάχιστον από τα 2 αυτά μπλοκ είναι τα $v_1$, $v_2$, $v_4$, $v_5$ και $v_6$, και αποτελούν το σύνολο $\mathcal{X}_{v_2}$.

Το αντίστοιχο μητρώο $\mathcal{D}$ μαζί με τα μητρώα $\mathbf{X}, \mathbf{Y}$ είναι τα ακόλουθα:

Η διαδικασία για τον υπολογισμό δυο γραμμών του μητρώου $\mathbf{D}$, $\mathbf{d}_{v_2}$ και $\mathbf{d}_{v_6}$, απεικονίζεται στο Σχήμα 3.2. Παρατηρούμε λοιπόν πως για να βρούμε το στοιχείο $d_{25}$ κοιτάμε σε ποια μπλοκ ανήκει το αντικείμενο $v_2$ και βλέπουμε πως ανήκει σε 2 μπλοκ, στο $\mathcal{D}_1$ και στο $\mathcal{D}_2$. Για την περίπτωση λοιπόν αυτή, το $N_{v_2} = 2$. Στη συνέχεια κοιτάμε σε ποιο μπλοκ

$$\mathbf{D} = \begin{pmatrix} 1 & 0 & 0 \\ \frac{1}{2} & \frac{1}{2} & 0 \\ 0 & 0 & 1 \\ \frac{1}{2} & \frac{1}{2} & 0 \\ 0 & 1 & 0 \\ \frac{1}{2} & 0 & \frac{1}{2} \end{pmatrix} \times \begin{pmatrix} \frac{1}{4} & \frac{1}{4} & 0 & \frac{1}{4} & 0 & \frac{1}{4} \\ 0 & \frac{1}{3} & 0 & \frac{1}{3} & \frac{1}{3} & 0 \\ 0 & 0 & \frac{1}{2} & 0 & 0 & \frac{1}{2} \end{pmatrix}$$

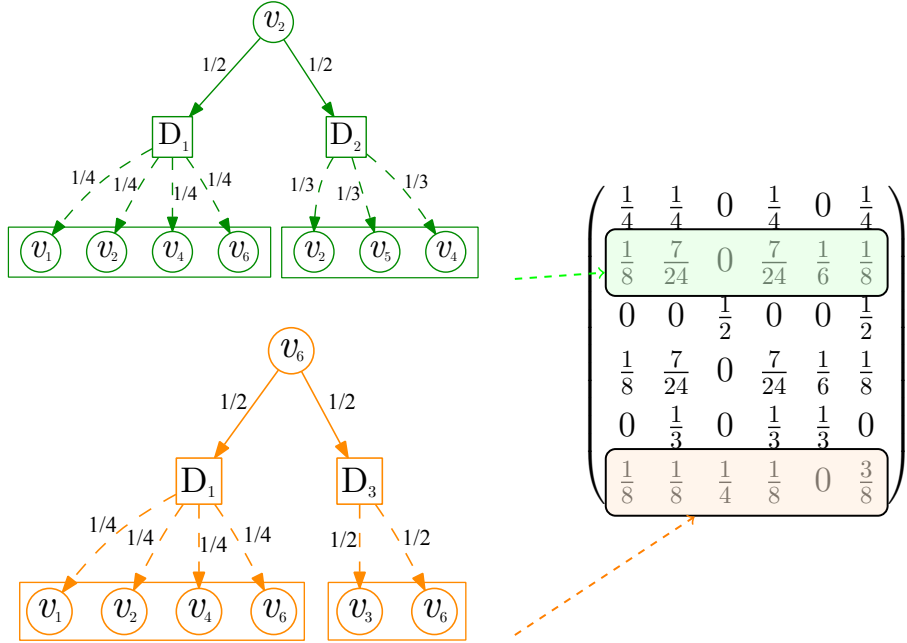

Σχήμα 3.2: Απεικόνιση του υπολογισμού των γραμμών $\mathbf{d}_{v_2}$ και $\mathbf{d}_{v_6}$.

ανήκει το αντικείμενο $v_5$ και βλέπουμε ότι ανήκει μόνο στο $\mathcal{D}_2$. Από τη στιγμή που στο μπλοκ $\mathcal{D}_2$ ανήκουν τρία αντικείμενα, το ποσοστό συσχέτισης που έχει πάρει το μπλοκ αυτό από το $v_2$ το μοιράζει ισοπίθανα σε αυτά. Συνεπώς, η τιμή του στοιχείου $d_{25}$ είναι $\frac{1}{6}$.

**Μητρώο προτίμησης χρήστη** Τέλος, όσον αφορά το *μητρώο προτίμησης* ενός χρήστη $u_i \in \mathcal{U}$, χρησιμοποιούμε το μητρώο τάξης-1 $\mathbf{E} \triangleq \mathbf{e}\boldsymbol{\omega}^{\mathsf{T}}$, όπου το $\mathbf{e}$ είναι ένα κατάλληλου μεγέθους διάνυσμα που όλα του τα στοιχεία είναι ίσα με 1 και το $\boldsymbol{\omega}^{\mathsf{T}}$ είναι το κανονικοποιημένο διάνυσμα των βαθμολογιών που έχει δώσει ο χρήστης για όλα τα αντικείμενα.

Από όσα αναφέρθηκαν παραπάνω, συμπεραίνουμε πως το *συνολικό εξατομικευμένο στοχαστικό μητρώο* $\mathbf{B}$ που προκύπτει από τα HIR μητρώα, προτείνει ένα μέτρο συσχέτισης των αντικειμένων. Το διάνυσμα στάσιμης κατανομής το οποίο αποτελεί την έξοδο του HIR αλγορίθμου, αποτελεί τη λίστα κατάταξης των αντικειμένων που θα προταθούν στο χρήστη. Στην επόμενη ενότητα ακολουθεί ο HIR αλγόριθμος.

## 3.4 Ο αλγόριθμος

Στο σημείο αυτό, ορίζουμε το **εξατομικευμένο διάνυσμα σύστασης**, το οποίο παράγεται από τον HIR, ως το *διάνυσμα στάσιμης κατανομής* $\boldsymbol{\pi}$ της αλυσίδας Markov το οποίο αντιστοιχεί στο στοχαστικό μητρώο, $\mathbf{B}$, με αρχική κατανομή $\boldsymbol{\omega}$. Ο HIR αλγόριθμος παρουσιάζεται παρακάτω.

---
**Algorithm 1** HIR $(\mathbf{C}, \mathbf{D}, \boldsymbol{\omega}, \texttt{tol}, \text{maxit})$

1. Έστω ότι η αρχική προσέγγιση είναι $\boldsymbol{\pi} = \boldsymbol{\omega}$. Θέσε $k = 0$.
2. Υπολόγισε

$$\begin{aligned}\hat{\boldsymbol{\pi}}^\mathsf{T} &:= \boldsymbol{\pi}^\mathsf{T}\mathbf{B} \\ &= \alpha\boldsymbol{\pi}^\mathsf{T}\mathbf{C} + \beta\boldsymbol{\pi}^\mathsf{T}\mathbf{D} + (1-\alpha-\beta)\boldsymbol{\pi}^\mathsf{T}\boldsymbol{\omega}\mathbf{e}^\mathsf{T}\end{aligned} \qquad (3.8)$$

3. Κανονικοποίησε το $\boldsymbol{\pi}^\mathsf{T}$ και υπολόγισε

$$r := \|\hat{\boldsymbol{\pi}}^\mathsf{T} - \boldsymbol{\pi}^\mathsf{T}\|_1$$

αν το $r < \texttt{tol}$ ή το $k \geq \text{maxit}$, τερμάτισε τον αλγόριθμο με $\hat{\boldsymbol{\pi}}^\mathsf{T}$, αλλιώς θέσε το $k := k+1$ και το $\boldsymbol{\pi} := \hat{\boldsymbol{\pi}}$
πήγαινε στο βήμα 2.

---

Προκειμένου να βρεθεί η *στάσιμη κατανομή*, εφαρμόζουμε τη *δυναμομέθοδο* [75] η οποία είναι ιδιαίτερα βολική για μεγάλα αραιά μητρώα. Όπως φαίνεται και από τα βήματα του αλγορίθμου, γίνεται αρχικοποίηση του διανύσματος της στάσιμης κατανομής με το ήδη γνωστό διάνυσμα προτίμησης του χρήστη, και στη συνέχεια εφαρμόζεται η μέθοδος μέχρι να συγκλίνει και να επιστρέψει τη λίστα κατάταξης του εκάστοτε χρήστη.

Ο αλγόριθμός μας, είναι πολύ αποδοτικός και από υπολογιστικής πλευράς και από πλευρά αποθήκευσης. Ο HIR χρειάζεται να αποθηκεύσει τα άμεσα μητρώα και το μητρώο NCD εγγύτητας. Το μητρώο $\mathbf{C}$ είναι εγγενώς αραιό και προσαρμόζεται πολύ καλά στην αύξηση του αριθμού των χρηστών. Η πρόσθεση ενός νέου χρήστη στο σύστημα θα μπορούσε να προκαλέσει μόνο την αύξηση των μη μηδενικών στοιχείων του $\mathbf{C}$, εφόσον η διάσταση του μητρώου εξαρτάται μόνο από την πληθικότητα του χώρου των αντικειμένων, ο οποίος είναι γεγονός πως στις περισσότερες πραγματικές εφαρμογές αυξάνει πολύ αργά. Στην περίπτωση του μητρώου $\mathbf{D}$, αν εκμεταλλευτούμε την παραγοντοποίηση που ορίζεται στη σχέση (3.7) και λάβουμε υπόψη μας το γεγονός ότι για οποιοδήποτε λογικό decom-position του χώρου αντικειμένων, το decomposition παραμένει $K \ll m$, δηλαδή τα μπλοκ παραμένουν λιγότερα από τα αντικείμενα, η αποθήκευση χρειάζεται να γίνει πολύ μικρή.

Επιπρόσθετα, από υπολογιστικής άποψης, το πιο "βαρύ" μέρος σε κάθε επανάληψη στον αλγόριθμό μας είναι ο αραιός Μητρώο×Διάνυσμα πολλαπλασιασμός $\boldsymbol{\pi}^{\mathsf{T}}\mathbf{C}$, ο οποίος συνεπάγεται μόλις $\Omega = 2 \times \mathrm{nnz}(\mathbf{C})$ λειτουργίες. Η παραγοντοποίηση του μητρώου NCD εγγύτητας και η υπερβολική αραιότητα των παραγόντων, επιβεβαιώνουν ότι το επιπρόσθετο βάρος το οποίο προκαλείται από το $\mathbf{D}$ είναι πολύ μικρό, και αν λάβουμε υπόψη μας το γεγονός ότι μπορεί να υπολογιστεί με έναν εντελώς παράλληλο τρόπο, το βάρος αυτό γίνεται σχεδόν αμελητέο. Οι ιδιότητες αυτές είναι πολύ χρήσιμες γιατί επιτρέπουν σε πραγματικά σενάρια την εισαγωγή περισσότερων από μια decompositions, σύμφωνα με διάφορα κριτήρια όπως ειπώθηκε προηγουμένως, πράγμα το οποίο οδηγεί σε καλύτερες συστάσεις. Στο δικό μας μοντέλο, η γενίκευση αυτή είναι πολύ εύκολη. Ο ορισμός περισσότερων συνόλων οικογενειών $\mathcal{D}^{(1)}, \mathcal{D}^{(2)}, \ldots$ και των αντίστοιχων NCD proximal συνόλων, οδηγεί στην εισαγωγή περισσοτέρων μητρώων NCD εγγύτητας $\mathbf{D}_1, \mathbf{D}_2, \ldots$ και σχετιζόμενων παραμέτρων $\beta_1, \beta_2, \ldots$ με ένα απλό τρόπο.

Επιπλέον, για λόγους απόδοσης, στην υλοποίησή μας χρησιμοποιούμε μια *batch* προσέγγιση της δυναμομεθόδου, δηλαδή μια "τροποποιημένη" έκδοση, η οποία υπολογίζει το διάνυσμα κατάταξης για όλους τους χρήστες μαζί. Ακολούθως, παραθέτουμε μέρος του κώδικα μας που υλοποιεί την batch προσέγγιση:

```
Pref = sparse(Pref);
PI=zeros(m,n)+full(Pref);
flag=0;
delta=2;
iter=0;
reshist=zeros(maxit,1);
PInew=zeros(m,n);

while iter<maxit && delta>tol

    PInew=a*(H'*PI);
    PInew=PInew + b*(C'*PI);
    PInew=PInew + c*(Sim'*PI);
    W = diag(sparse(ones(1,n) - sum(PInew)));
    PInew = PInew + Pref*W;
    D_PI = PI-PInew;  delta=norm(D_PI,1);
    reshist(iter+1)=delta;
    iter=iter+1;
    PI=PInew*diag(sparse(1./sum(PInew)));
end

flag=delta>tol;  reshist=reshist(1:iter);
```

```
fprintf('Solved BatchHIR 2 (a=%6.5f, b=%6.5f, c=%6.5f) in %5i multiplies to
    %8e tolerance\n', a, b, c, iter, delta);
```

Listing 3.1: Κώδικας υπολογισμού HIR με αξιοποίηση υπορουτίνων BLAS-3

Το **Pref** μητρώο αποτελεί το μητρώο προτίμησης όλων των χρηστών και αποτελείται από τόσες *γραμμές* όσα είναι τα αντικείμενα του συστήματος και τόσες *στήλες* όσοι είναι οι χρήστες. Δηλαδή, η πρώτη στήλη του θα περιέχει το διάνυσμα προτίμησης του πρώτου χρήστη για όλα τα αντικείμενα. Το **PI** αποτελεί το μητρώο κατάταξης των αντικειμένων που θα επιστραφεί από τον HIR, όταν η batch προσέγγιση συγκλίνει. Αντίστοιχα με το μητρώο Pref, η πρώτη στήλη του PI θα περιέχει την επιστρεφόμενη λίστα κατάταξης του πρώτου χρήστη.

Για την πλήρη κατανόηση του κώδικα παραθέτουμε τα εξής:

- Η εντολή "**PI=zeros(m,n)+full(Pref);**" αρχικοποιεί το μητρώο κατάταξης των αντικειμένων με τις προσωπικές προτιμήσεις όλων των χρηστών.

- Με τις εντολές "**W = diag(sparse(ones(1,n)) - sum(PInew)));**" και "**PInew = PInew + Pref*W;**" προσθέτουμε το μητρώο που περιέχει όλα τα κανονικοποιημένα διανύσματα προτίμησης των χρηστών, το οποίο αποτελεί και το τρίτο συστατικό του HIR.

Η επιλογή της τροποποιημένης έκδοσης της δυναμομεθόδου έγινε με σκοπό να χρησιμοποιήσουμε τις BLAS-3 (πράξεις μεταξύ μητρώων) προκειμένου να εκμεταλλευτούμε την ιεραρχία μνήμης του συστήματός μας. Για την επιβεβαίωση της θετικής επίδρασης της τροποποιημένης δυναμομεθόδου, τρέχουμε 100 φορές τον αλγόριθμό μας και για τις δυο περιπτώσεις.

Στην μεν πρώτη περίπτωση - όπου χρησιμοποιούμε την κλασσική δυναμομέθοδο - τρέχουμε 100 φορές τον HIR για κάθε ένα χρήστη ξεχωριστά, ενώ στη δεύτερη - με την batch προσέγγιση - τον τρέχουμε 100 φορές για όλους τους χρήστες μαζί. Τα αποτελέσματα του πειράματος φαίνονται στο Σχήμα 3.3. Οι προδιαγραφές του συστήματος που γίνονται τα πειράματα βρίσκονται στο Παράρτημα Β.

Είναι εμφανές πως με τη batch προσέγγιση ο HIR είναι πιο αποδοτικός. Συγκεκριμένα, ο χρόνος που χρειάζεται να τρέξει για να παράξει τη λίστα κατάταξης των προτεινόμενων αντικειμένων για όλους τους χρήστες είναι κατά μέσο όρο ίσος με 4,75 δευτερόλεπτα. Ενώ με την απλή δυναμομέθοδο ο χρόνος που χρειάζεται είναι μεγαλύτερος και ίσος κατά μέσο όρο με 39,09 δευτερόλεπτα, δηλαδή χρειάζεται περίπου 0,041 δευτερόλεπτα για να συγκλίνει για ένα μόνο χρήστη. Σε πραγματικές καταστάσεις, όπου υπάρχουν περισσότερα αντικείμενα και περισσότεροι χρήστες, η απαίτηση αυτή είναι πολύ πιο σημαντική.

Έπειτα από την αναλυτική περιγραφή του μοντέλου μας, θα συνεχίσουμε στο επόμενο κεφάλαιο με τη διεξαγωγή των πειραμάτων πάνω σε συγκεκριμένο dataset, προκειμένου να

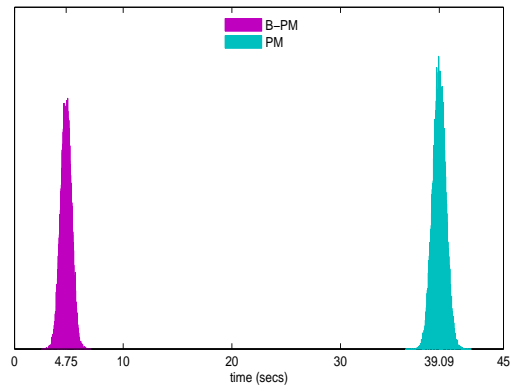

Σχήμα 3.3: Σύγκριση μεταξύ της "batch" και της κλασσικής δυναμομεθόδου .

αξιολογήσουμε την απόδοσή του με βάση κάποιες μετρικές απόδοσης.

# Κεφάλαιο 4

# Πειραματική αξιολόγηση

Στο Κεφάλαιο 4 εφαρμόζουμε το HIR framework στο *movie recommendation* πρόβλημα και παραθέτουμε μια σειρά από πειράματα προκειμένου να μετρήσουμε την απόδοση της μεθόδου μας. Τα πειράματα αυτά πραγματοποιούνται πάνω στο MovieLens dataset [1]. Αρχικά, η απόδοση σύστασης του HIR συγκρίνεται με τα αποτελέσματα άλλων αλγορίθμων κατάταξης, των οποίων οι τιμές βρίσκονται στη βιβλιογραφία.

Στη συνέχεια, προσεγγίζουμε το πρόβλημα της αραιότητας του χώρου των αντικειμένων από δυο διαφορετικές όψεις. Συγκεκριμένα, στόχος μας είναι να δούμε πώς συμπεριφέρεται ο HIR αλγόριθμος στην περίπτωση που δεν υπάρχουν πολλές διαθέσιμες βαθμολογίες ταινιών αλλά και κατά την εισαγωγή νέων ταινιών στο σύστημα. Για τη μέτρηση της απόδοσης χρησιμοποιούμε πολύ γνωστές μετρικές απόδοσης, όπως είναι το Degree Of Agreement και ο Kendall's $\tau$ συντελεστής συσχέτισης.

## 4.1 MovieLens dataset

Το `MovieLens` dataset είναι μια δημοφιλής βάση δεδομένων που δημιουργήθηκε από την ομάδα έρευνας GroupLens του πανεπιστημίου της Minnesota. Αυτή η βάση δεδομένων, η οποία περιέχει δεδομένα που έχουν συλλεχθεί από ένα δημοφιλές σύστημα συστάσεων ταινιών, έχει αξιοποιηθεί ευρέως σαν benchmark για την αξιολόγηση των διαφόρων προτεινόμενων προσεγγίσεων, που υπάρχουν στη βιβλιογραφία, για τη δημιουργία συστημάτων συστάσεων. Περιέχει 100,000 βαθμολογίες από 943 χρήστες για 1,682 ταινίες. Κάθε χρήστης έχει βαθμολογήσει τουλάχιστον 20 ταινίες, προκειμένου να υπολογιστεί με μεγαλύτερη αξιοπιστία το προφίλ του. Κάθε βαθμολογία αναπαρίσταται παίρνει μια ακέραια τιμή ανάμεσα στο 1 και στο 5.

Το `MovieLens` dataset περιλαμβάνει 5 προκαθορισμένα splittings, δηλαδή τα δεδομένα του dataset έχουν χωριστεί σε 5 διαφορετικά training και test set αντίστοιχα, για καθένα από τα οποία ισχύει πώς το 80% των βαθμολογιών τους αποτελεί το *training set* $\mathcal{L}$, με βάση το οποίο θα "μάθει" ο αλγόριθμός μας και το 20% το *test set* $\mathcal{T}$ (όπως περιγράφεται στο [70]) το οποίο χρησιμοποιείται κατά τη διαδικασία ελέγχου της απόδοσής του.

---

[1] http://www.movielens.umn.edu

Επιπλέον ορίζουμε, το $\mathcal{W}_i \triangleq \mathcal{L}_i \cup \mathcal{T}_i$, σαν το σύνολο των ταινιών που έχει δει ο χρήστης $u_i$. Συνεπώς, το σύνολο $\overline{\mathcal{W}_i} \subset \mathcal{V}$ αφορά ταινίες που δεν υπάρχουν ούτε στο training set ούτε στο test set (δηλαδή, ο χρήστης δεν έχει δει αυτές τις ταινίες).

Το MovieLens dataset περιλαμβάνει επίσης πληροφορίες που αφορούν και τις κατηγορίες στις οποίες ανήκει κάθε ταινία. Αυτό είναι και το κριτήριο του decomposition, όπως το αναφέρουμε στην Ενότητα 3.2, που χρησιμοποιούμε για την πειραματική αξιολόγηση της μεθόδους μας. Αξίζει να σημειωθεί, για ακόμα μια φορά, πώς σε πραγματικές καταστάσεις, όπου υπάρχει περισσότερη διαθέσιμη πληροφορία για τα δεδομένα, είναι δυνατόν να υπάρξουν περισσότερες από μια decompositions του χώρου των αντικειμένων.

## 4.2 Μετρικές απόδοσης

### 4.2.1 Degree of agreement

Η μετρική *DOA* χρησιμοποιείται στη βιβλιογραφία για την αξιολόγηση της ποιότητας των ranking-based μεθόδων σύστασης. Για τον υπολογισμό του DOA, λοιπόν, το οποίο μετράει πόσο ακριβής είναι η λίστα κατάταξης των ταινιών για ένα χρήστη, ακολουθείται η παρακάτω διαδικασία [20].

Έστω ένα ζευγάρι ταινιών, όπου η μια ταινία (έστω $v_1$) ανήκει στο test set για ένα συγκεκριμένο χρήστη (ο χρήστης αυτός, δηλαδή, έχει ήδη δει την ταινία) και η άλλη ταινία (έστω $v_2$) δεν ανήκει ούτε στο test set αλλά ούτε και στο training set (ο χρήστης, δηλαδή, δεν έχει δει αυτή την ταινία). Η διαδικασία αυτή επαναλαμβάνεται για κάθε ένα από τα ζευγάρια ταινιών που υπάρχουν στο dataset. Ο αλγόριθμός μας, λοιπόν, ταξινομεί τα ζευγάρια στη σωστή σειρά αν, στη λίστα κατάταξης η οποία υπολογίστηκε από το training set, η ταινία $v_1$ προηγείται της ταινίας $v_2$. Το ατομικό DOA είναι συνεπώς το ποσοστό των ζευγαριών τα οποία ταξινομήθηκαν σε σωστή σειρά στη λίστα δια το συνολικό αριθμό των ζευγαριών. Η ιδέα πίσω από αυτή τη μετρική είναι πως ένας καλός αλγόριθμος συστάσεων οφείλει να προτιμήσει τις ταινίες τις οποίες ο χρήστης πράγματι έχει δει, με το να τις ταξινομήσει σε υψηλότερο επίπεδο από ότι ταξινομεί εκείνες που δεν έχει δει καθόλου. Μια τυχαία κατάταξη παράγει ένα ποσοστό 50% του degree of agreement, το οποίο σημαίνει ότι το 50% των συνολικών ζευγαριών ταξινομήθηκαν στη σωστή σειρά και το υπόλοιπο 50% στη λάθος. Ένα ποσοστό 100% του degree of agreement σημαίνει πώς η προτεινόμενη κατάταξη είναι πανομοιότυπη με την ιδανική κατάταξη.

Σύμφωνα λοιπόν με τα παραπάνω η μετρική απόδοσης DOA ορίζεται ως ακολούθως:

$$\text{DOA}_i \triangleq \frac{\sum_{v_j \in \mathcal{T}_i \land v_k \in \overline{\mathcal{W}}_i} \left[ \pi^i(v_j) > \pi^i(v_k) \right]}{\mid \mathcal{T}_i \mid * \mid \overline{\mathcal{W}}_i \mid}$$

όπου το $\pi^i(v_j)$ είναι το σκορ κατάταξης της ταινίας $v_j$ στη λίστα κατάταξης του χρήστη $u_i$, και το $[S]$ ισούται με 1, αν η πρόταση $S$ είναι αληθής, αλλιώς είναι 0 [1]. Εφόσον ορίσαμε το DOA σχετικά με τον κάθε χρήστη $u_i$, προσδιορίζουμε στη συνέχεια το *macro-averaged DOA* (macro-DOA) και το *micro-averaged DOA* (micro-DOA).

Συγκεκριμένα, το macro-DOA είναι ο μέσος όρος όλων των ατομικών $\text{DOA}_i$, όπου το $\mathcal{T}_i$ δεν είναι κενό, και δίνεται από τον τύπο:

$$\text{macro-DOA} \triangleq \frac{\sum_{\mathcal{T}_i \neq 0} \text{DOA}_i}{|\{\mathcal{T}_j : \mathcal{T}_j \neq \emptyset\}|}$$

και το micro-DOA είναι ο λόγος μεταξύ των συνολικού αριθμού των ζευγαριών των ταινιών που είναι στη σωστή σειρά και του συνολικού αριθμού των ζευγαριών τα οποία έχουν ελεχθεί [20, 23], και δίνεται από τον τύπο:

$$\text{micro-DOA} \triangleq \frac{\sum_{\mathcal{T}_i \neq 0} \sum_{v_j \in \mathcal{T}_i \land v_k \in \overline{\mathcal{W}}_i} \left[ \pi^i(v_j) > \pi^i(v_k) \right]}{\sum_{\mathcal{T}_i \neq 0} \mid \mathcal{T}_i \mid * \mid \overline{\mathcal{W}}_i \mid}$$

Είναι προφανές πώς, το micro-DOA αναθέτει ένα υψηλότερο βάρος στους χρήστες των οποίων οι βαθμολογίες στο test set είναι πολλές, ενώ το macro-DOA αναθέτει το ίδιο βάρος σε όλους τους χρήστες, χωρίς να έχει σημασία πόσες πολλές βαθμολογίες υπάρχουν στο test set για καθέναν από αυτούς.

### 4.2.2 Ο Kendall's $\tau$ συντελεστής συσχέτισης

Ο Kendall's $\tau$ συντελεστής συσχέτισης, είναι ένα μέτρο που δείχνει τη συσχέτιση μεταξύ δυο μεταβλητών. Πρόκειται για ένα διαισθητικό μη παραμετρικό δείκτη συσχέτισης ο οποίος έχει χρησιμοποιηθεί ευρέως τα τελευταία χρόνια στην κοινότητα του Web (για παράδειγμα [3, 36]) για συγκρίσεις λιστών κατάταξης. Ο Kendall's $\tau$ συντελεστής συσχέτισης ορίζεται ως ακολούθως:

$$\tau = \frac{(\# \text{ ζευγαριών στη σωστή σειρά}) - (\# \text{ ζευγαριών στη λάθος σειρά})}{\frac{1}{2}n(n-1)}$$

Δυο ζευγάρια ταινιών, έστω $(x_i, y_i)$ και $(x_j, y_j)$, λέμε ότι είναι στη σωστή σειρά αν οι ταξινομήσεις στη λίστα κατάταξης και για τα δυο στοιχεία συμφωνούν, δηλαδή αν $x_i > x_j$ και $y_i > y_j$ ή $x_i < x_j$ και $y_i < y_j$. Δυο τέτοια ζευγάρια είναι στη λάθος σειρά, αν $x_i > x_j$ και $y_i < y_j$ ή αν $x_i < x_j$ και $y_i > y_j$. Αν $x_i = x_j$ και $y_i = y_j$, τότε τα ζευγάρια δεν είναι ούτε στη σωστή ούτε στη λάθος σειρά. Όσον αφορά τις ιδιότητες του Kendall's $\tau$ ισχύουν τα εξής:

- Ο παρονομαστής είναι ο συνολικός αριθμός των ζευγαριών, συνεπώς ο συντελεστής πρέπει να έχει τιμή ανάμεσα στο διάστημα $-1 \leqslant \tau \leqslant 1$.

- Αν η συμφωνία ανάμεσα στις δυο λίστες κατάταξης είναι τέλεια (δηλαδή οι δυο κατατάξεις είναι ίδιες), ο συντελεστής έχει τιμή ίση με 1.

- Αν η διαφωνία ανάμεσα στις δυο λίστες κατάταξης είναι τέλεια (δηλαδή η μια κατάταξη είναι το αντίθετο της άλλης), ο συντελεστής έχει τιμή ίση με -1.

- Αν η **x** και η **y** είναι ανεξάρτητες, τότε θα περιμένουμε ο συντελεστής να είναι περίπου μηδέν.

---

[1] αυτή η χρήσιμη σημειογραφία αναφέρεται στη βιβλιογραφία σαν Iverson σημειογραφία, βλέπε [28].

## 4.3 Διαδικασία πειραμάτων και αποτελέσματα

Στο σημείο αυτό, είμαστε σε θέση να προχωρήσουμε στη διεξαγωγή των πειραμάτων καθώς και στην ανάλυση των αποτελεσμάτων της μεθόδου μας. Προηγουμένως όμως θα αναφέρουμε λίγα λόγια για τους αλγορίθμους με τους οποίους συγκρινόμαστε.

### 4.3.1 Σύγκριση του HIR με άλλους rank-based αλγορίθμους συστάσεων

Αρχικά, ο *ItemRank* (IR), ο οποίος περιγράφτηκε εν συντομία στην Ενότητα 2.3.2, προτάθηκε από τους Gori and Pucci [27]. Οφείλει την έμπνευσή του στον PageRank και παράγει ένα εξατομικευμένο διάνυσμα κατάταξης χρησιμοποιώντας το γράφο συσχέτισης των αντικειμένων. Ο αλγόριθμος TR ο οποίος προτάθηκε από τους Zhang and Li [85] αποτελεί μια βελτίωση του ItemRank και βασίζεται σε μια topical εκδοχή του PageRank [52]. Για τον υπολογισμό της λίστας κατάταξης των αντικειμένων λαμβάνει, επιπλέον, υπόψη του την κατηγορία του αντικειμένου και το προφίλ ενδιαφέροντος του χρήστη.

Οι Fouss et al. [22] ακολουθώντας μια γραφική αναπαράσταση των δεδομένων, και χρησιμοποιώντας μια προσέγγιση βασισμένη στους τυχαίους περιπάτους, παρουσίασαν έναν αριθμό από μεθόδους - για τον υπολογισμό διαφόρων μέτρων ομοιότητας των κόμβων - όπως είναι τα *average commute time* (απλό CT and PCA-CT) και ο *ψευδοαντίστροφος του Laplacian μητρώου* **L** (**L**$^\dagger$). Συγκεκριμένα, για το average commute time (CT) το οποίο περιγράψαμε και στην Ενότητα 2.3.2, ισχύει πώς όσο μικρότερη είναι η απόσταση ανάμεσα σε ένα χρήστη και ένα αντικείμενο, τόσο το αντικείμενο αυτό βρίσκεται πιο ψηλά στην προτεινόμενη λίστα κατάταξης για τον χρήστη. Η τετραγωνική ρίζα του average commute time είναι μια Ευκλείδεια απόσταση. Όσον αφορά τη βασισμένη στην *Ευκλείδεια commute time απόσταση PCA τεχνική* (PCA CT), ισχύει πώς από την ανάλυση σε ιδιοδιανύσματα του $\mathbf{L}^\dagger$, είναι δυνατόν να γίνει η αντιστοίχηση των κόμβων του γράφου σε ένα νέο Ευκλείδειο χώρο, ο οποίος διατηρεί την Ευκλείδεια commute time απόσταση (ECTD). Είναι επίσης πιθανό, οι κόμβοι να προβληθούν σε ένα $M$-διάστατο υποχώρο, εφαρμόζοντας τη PCA τεχνική και κρατώντας ένα δεδομένο αριθμό από κύρια χαρακτηριστικά. Στη συνέχεια οι αποστάσεις, οι οποίες υπολογίζονται μεταξύ των κόμβων στο χώρο των μειωμένων διαστάσεων, μπορούν να χρησιμοποιηθούν για την ταξινόμηση των αντικειμένων για κάθε χρήστη. Το $M$ (δηλαδή η διάσταση του υποχώρου), είναι ο αριθμός των κύριων συνιστωσών (principal components). Επιπρόσθετα, το $\mathbf{L}^\dagger$ παρέχει ένα μέτρο ομοιότητας $(sim(i,j) = l_{ij}^\dagger)$ εφόσον είναι το μητρώο το οποίο περιέχει τα εσωτερικά γινόμενα των διανυσμάτων στον Ευκλείδειο χώρο όπου οι κόμβοι διαχωρίζονται μέσω του ECTD. Εφόσον λοιπόν υπολογιστεί το μητρώο ομοιότητας, τα αντικείμενα ταξινομούνται με βάση τις ομοιότητες τους με τον χρήστη.

Επιπρόσθετα, ο *Katz* [38] πρότεινε μια μέθοδο υπολογισμού ομοιοτήτων λαμβάνοντας υπόψη και τον αριθμό των άμεσων αλλά και των έμμεσων συνδέσεων μεταξύ των αντικειμένων[1]. Ο Maximum-Frequency (MaxF) αλγόριθμος κατάταξης, ταξινομεί απλά τα αντικείμενα με βάση τον αριθμό των ατόμων που τα έχουν δει. Με άλλα λόγια, τα αντικείμενα προτείνονται σε κάθε χρήστη με φθίνουσα δημοτικότητα. Συνεπώς, η λίστα κατάταξης των αντικειμένων θα είναι ίδια για όλους τους χρήστες, αφού δεν εξαρτάται από τις μέχρι τώρα προτιμήσεις τους. Ο συγκεκριμένος αλγόριθμος μας εξυπηρετεί σαν αναφορά για την εκτίμηση της ποιότητα των άλλων αλγορίθμων κατάταξης.

Τέλος, οι Freno et al. [23] πρότειναν ένα Hybrid Random Fields μοντέλο (HRF) το οποίο εφάρμοσαν, μαζί με έναν αριθμό από πολύ γνωστά πιθανοτικά γραφικά μοντέλα, όπως είναι τα Depedency Networks (DN), Markov Random Fields (MRF) και Naive Bayes (NB), για να προβλέψουν τα κορυφαία Ν αντικείμενα για τους χρήστες. Συγκεκριμένα, για την πρόβλεψη του ενδιαφέροντος ενός χρήστη $u_i$ για ένα αντικείμενο $v_j$, τα μοντέλα αυτά υπολογίζουν την υπό συνθήκη πιθανότητα ο χρήστης να διαλέξει το αντικείμενο $v_j$, το οποίο ακόμα δεν έχει δει, με βάση τα αντικείμενα τα οποία έχει ήδη επιλέξει. Η τιμή λοιπόν της υπολογισμένης υπό συνθήκη πιθανότητας, χρησιμοποιείται σαν την τιμή του ενδιαφέροντος του χρήστη $u_i$ για το αντικείμενο $v_j$. Η προβλεπόμενη λίστα προτίμησης των χρηστών προκύπτει από τη σύγκριση των υπό συνθήκη πιθανοτήτων όλων των αντικειμένων.

**Αποτελέσματα** Για τη σύγκριση της μεθόδου μας με τους παραπάνω αλγορίθμους κατάταξης, χρησιμοποιούμε τις μετρικές micro-DOA και macro-DOA οι οποίες υπολογί-

---

[1] Περισσότερες λεπτομέρειες για το μέτρο Katz, παραθέσαμε στην Ενότητα 2.3.1.

| micro-DOA | | macro-DOA | | | |
|---|---|---|---|---|---|
| DN [18] | $81.33 \pm 0.43$ | DN | $80.51 \pm 1.23$ | MRF | $89.47 \pm 0.44$ |
| HRF [23] | $88.07 \pm 0.59$ | HRF | $89.83 \pm 0.52$ | NB | $88.87 \pm 0.22$ |
| IR[27] | $87.06 \pm 0.10$ | IR | $87.76 \pm 0.27$ | **HIR** | $\mathbf{89.99 \pm 0.20}$ |
| **HIR** | $\mathbf{88.85 \pm 0.29}$ | Katz [20, 38] | $85.83 \pm 0.24$ | CT[20] | $84.09 \pm 0.01$ |
| MRF[23] | $88.09 \pm 0.50$ | $L^{\dagger}$ [20] | $87.23 \pm 0.84$ | PCA CT [20] | $84.04 \pm 0.76$ |
| NB[18, 23] | $86.66 \pm 0.30$ | MaxF[20] | $84.07 \pm 0.00$ | TR[85] | $89.08 \pm 0.11$ |

Πίνακας 4.1: Σύγκριση απόδοσης μεταξύ του HIR και πολλών άλλων state-of-the-art αλγόριθμων σύστασης, χρησιμοποιώντας τις μετρικές micro-DOA και macro-DOA.

στηκαν με 5-fold cross-validation. Όπως αναφέρθηκε προηγουμένως, για τη διεξαγωγή των πειραμάτων χρησιμοποιούμε το MovieLens dataset με τα προκαθορισμένα πέντε splittings, προκειμένου να μας επιτραπεί να συγκριθούμε με τα διαφορετικά αποτελέσματα, των παραπάνω αλγορίθμων, που βρίσκονται στη βιβλιογραφία. Τα αποτελέσματα της σύγκρισης παρουσιάζονται στον Πίνακα 4.1.

Παρατηρώντας, λοιπόν, τα αποτελέσματα του πίνακα βλέπουμε πως ο HIR ξεπερνά τις άλλες state-of-the-art τεχνικές, πετυχαίνοντας μια τιμή για το macro-DOA ίση με **89.99** η οποία είναι περίπου 6.8% μεγαλύτερη από την τιμή που πετυχαίνει ο MaxF τον οποίο όπως αναφέραμε παραπάνω τον χρησιμοποιούμε σαν αναφορά για την εκτίμηση της ποιότητας των άλλων αλγορίθμων κατάταξης. Το ίδιο ισχύει και για το μέτρο micro-DOA, όπου στην περίπτωση αυτή ο HIR πετυχαίνει την τιμή **88.85** σε αντίθεση με τους MRF και HRF όπου πετυχαίνουν τις τιμές 88.09 και 88.07 αντίστοιχα. Επιπλέον, παρατηρούμε πως ο HIR έχει και καλύτερη τυπική απόκλιση σε σχέση με τους άλλους δυο. Πολύ καλή τυπική απόκλιση παρουσιάζει, επίσης, και στην περίπτωση του macro-DOA όπως φαίνεται και από τον πίνακα. Όσον αφορά τον MaxF η τυπική απόκλιση είναι μηδέν, πράγμα λογικό αφού σε όλες τις περιπτώσεις η λίστα κατάταξης των ταινιών είναι ίδια για όλους τους χρήστες.

### 4.3.2  Πρόβλημα αραιών δεδομένων

Κύριο μέλημα της ενότητας αυτής είναι να δούμε πως συμπεριφέρεται ο HIR αλγόριθμος όταν ο χώρος αντικειμένων είναι αραιός. Είναι γεγονός πως το πρόβλημα της αραιότητας των δεδομένων αποτελεί ένα μείζον θέμα των συστημάτων συστάσεων. Εξαιτίας του γεγονότος αυτού, αναλύουμε το πρόβλημα από δυο διαφορετικές οπτικές γωνίες. Στην πρώτη περίπτωση, θεωρούμε πως δεν υπάρχουν πολλές βαθμολογίες ταινιών στο training set και στη δεύτερη ότι στο σύστημα εισάγονται καινούριες ταινίες (τοπική αραιότητα). Η μετρικές απόδοσης που χρησιμοποιούνται είναι το DOA και ο Kendall's $\tau$.

**Αραιότητα**  Προκειμένου να δείξουμε τα πλεονεκτήματα της μεθόδους μας απέναντι στα προβλήματα που προκαλούνται εξαιτίας της αραιότητας του χώρου αντικειμένων, διεξάγουμε το ακόλουθο πείραμα. Για καθένα από τα 5 προκαθορισμένα splittings, προσο-μοιώνουμε το φαινόμενο της αραιότητας του χώρου αντικειμένων διαλέγοντας με τυχαίο τρόπο το 80%, 60%, και 40% των βαθμολογιών των ταινιών τις οποίες συμπεριλαμβάνουμε σε τρεις τεχνητά αραιές εκδόσεις του. Στη συνέχεια χρησιμοποιούμε τις εκδόσεις αυτές για να υπολογίσουμε το πόσο επηρεάζεται η ποιότητα των συστάσεων, από την αραιότητα. Για να απομονώσουμε τη θετική επίδραση που έχει η εισαγωγή της NCD εγγύτητας και το σχετικό μητρώο **D**, τρέχουμε τον HIR καθώς και τους δυο αλγορίθμους που αποτελούν τα βασικά του υπο-συστατικά (τον ItemRank και τον SimRank[1]) και εκτιμούμε την απόδοσή τους τρέχοντας τα στάνταρ degree of agreement τεστ.

Στη γραφική παράσταση του Σχήματος 4.1, βλέπουμε πως ο HIR συμπεριφέρεται καλύτερα από τους άλλους δυο αλγορίθμους και στις δυο μετρικές micro-DOA και macro-DOA, παρουσιάζοντας πολύ καλά αποτελέσματα ακόμα και αν μόνο το 40% των βαθμολογιών είναι διαθέσιμο. Τα αποτελέσματα αυτά επιβεβαιώνουν τη διαίσθηση πίσω από τον HIR· ακόμα και αν οι άμεσες σχέσεις μεταξύ των ταινιών του dataset "καταρρέουν" από την αφαίρεση τόσων πολλών βαθμολογιών, η "μεταξύ επιπέδων" εγγύτητα που συλλαμβάνεται από το μητρώο **D** βάλλεται δυσκολότερα και ως εκ τούτου διαφυλάσσει την πιο αδρή δομή των δεδομένων. Αυτό οδηγεί σε ένα διάνυσμα κατάταξης το οποίο αποδεικνύεται λιγότερο ευαίσθητο στην αραιότητα του χώρου των αντικειμένων.

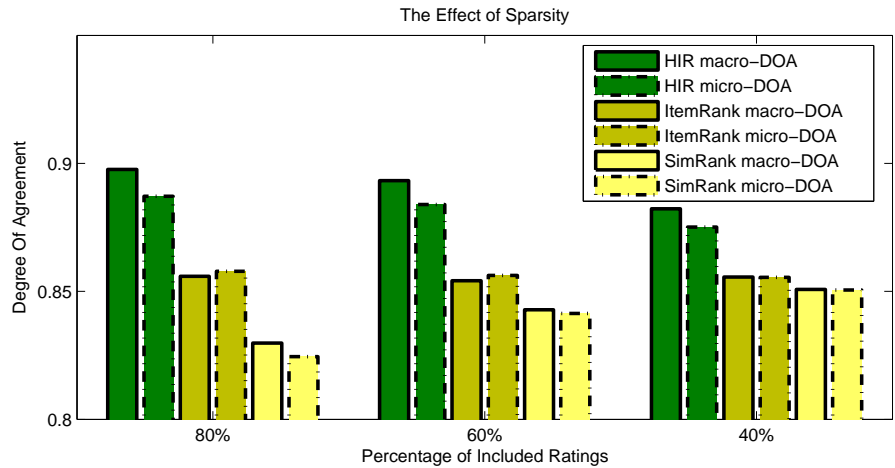

Σχήμα 4.1: Η επίδραση της αραιότητας στην ποιότητα των συστάσεων για τις τεχνητά αραιές εκδόσεις του `MovieLens` dataset.

**Τοπική αραιότητα**   Μια πολύ συχνή και ενδιαφέρουσα εμφάνιση του προβλήματος της αραιότητας προκύπτει μέσω της εισαγωγής νέων ταινιών στο σύστημα. Είναι φυσικό πως τα αντικείμενα αυτά θα έχουν βαθμολογηθεί πολύ λίγες φορές αφού είναι καινούρια, και συνεπώς η σχέση τους με τα υπόλοιπα στοιχεία του χώρου αντικειμένων δεν είναι ακόμα ξεκάθαρη. Αυτό, σε πολλές περιπτώσεις, θα μπορούσε να οδηγήσει σε άδικη μεταχείρισή τους.

---

[1] Ο SimRank είναι μια απλή παραλλαγή του ItemRank και συσχετίζει τα αντικείμενα χρησιμοποιώντας το adjusted cosine-based μητρώο ομοιότητας που ορίζεται στη σχέση 3.4, αντί για το μητρώο συσχέτισης που ορίζεται στη σχέση 3.3.

Για να αξιολογήσουμε λοιπόν την απόδοση του αλγορίθμου μας, έπειτα από την εισαγωγή νέων ταινιών, διεξάγουμε το ακόλουθο πείραμα. Αρχικά, επιλέγουμε τυχαία από το dataset έναν αριθμό από ταινίες οι οποίες έχουν βαθμολογηθεί τουλάχιστον 30 φορές και στη συνέχεια διαγράφουμε με τυχαίο τρόπο το 90% των βαθμολογιών τους. Η ιδέα είναι πως τα τροποποιημένα δεδομένα αναπαριστούν μια "*προηγούμενη έκδοση*" του dataset, όπου οι ταινίες αυτές είναι νέες στο σύστημα και συνεπώς έχουν λιγότερες βαθμολογίες.

Τρέχουμε διαφορετικές περιπτώσεις του HIR για διαφορετικές τιμές του $\beta$, του οποίου οι τιμές κυμαίνονται από $\beta = 0$ (όπου το μητρώο **D** δεν συμπεριλαμβάνεται και η λίστα κατάταξης των αντικειμένων παράγεται από τον σταθμισμένο συνδυασμό των άμεσων υποσυστατικών) μέχρι $\beta = 4$, κρατώντας το άθροισμα $\alpha + \beta = 0.9$. Στη συνέχεια, συγκρίνουμε τις κατατάξεις που προέρχονται από τα τροποποιημένα δεδομένα με τις αντίστοιχες αρχικές τους καταστάσεις. Το μέτρο που χρησιμοποιούμε για τη σύγκριση αυτή είναι ο συντελεστής συσχέτισης Kendall's $\tau$.

Υψηλή τιμή αυτής της μετρικής σημαίνει ότι οι δυο λίστες κατάταξης είναι πολύ "κοντά", το οποίο σημαίνει ότι οι νέες ταινίες που προστίθενται στο σύστημα είναι πολύ πιθανό να λάβουν παρόμοια μεταχείριση με αυτή που παίρνουν όταν συμπεριλαμβάνονται όλες οι διαθέσιμες βαθμολογίες. Τέλος, προκειμένου να έχουμε ένα μέτρο της ποιότητας των αρχικών λιστών κατάταξης για κάθε μια από τις περιπτώσεις του HIR, τρέχουμε τις μετρικές micro-DOA και macro-DOA και παρουσιάζουμε τα αποτελέσματα στον Πίνακα 4.2.

| # Νέων ταινιών | $\alpha = 0.9$ $\beta = 0$ | $\alpha = 0.85$ $\beta = 0.05$ | $\alpha = 0.8$ $\beta = 0.1$ | $\alpha = 0.7$ $\beta = 0.2$ | $\alpha = 0.6$ $\beta = 0.3$ | $\alpha = 0.5$ $\beta = 0.4$ |
|---|---|---|---|---|---|---|
| 100 | 0.8736 | 0.8776 | 0.8812 | 0.8878 | 0.8945 | 0.9020 |
| 200 | 0.7814 | 0.7886 | 0.7949 | 0.8066 | 0.8186 | 0.8317 |
| 300 | 0.6843 | 0.6940 | 0.7025 | 0.7190 | 0.7369 | 0.7572 |
| macro-DOA | $89.63 \pm 0.18$ | $89.91 \pm 0.20$ | **$89.99 \pm 0.20$** | $89.58 \pm 0.22$ | $88.45 \pm 0.23$ | $86.62 \pm 0.25$ |
| micro-DOA | $88.47 \pm 0.28$ | $88.75 \pm 0.29$ | **$88.85 \pm 0.29$** | $88.50 \pm 0.29$ | $87.42 \pm 0.29$ | $85.63 \pm 0.29$ |

Πίνακας 4.2: Απόδοση του HIR όσον αφορά την τοπική αραιότητα.

Τα πειράματα δείχνουν πως η εισαγωγή ακόμα και ενός πολύ μικρού $\beta$ επιφέρει μια θετική επίδραση στην περίπτωση της τοπικής αραιότητας. Αυτό έρχεται σε απόλυτη συμφωνία με τον τρόπο που ο HIR βλέπει το πρόβλημα. Στη μέθοδό μας, το σκορ κατάταξης ενός αντικειμένου δεν καθορίζεται αποκλειστικά από τις βαθμολογίες που έχει πάρει· συμβάλει και η κατηγορία στην οποία ανήκει γιατί σχετίζει το νέο αντικείμενο με το NCD proximal σύνολό του. Συνεπώς, ακόμα και αν οι βαθμολογίες δεν επαρκούν, οι νέες προστιθέμενες ταινίες παίρνουν τη θέση που τους αξίζει.

Όπως ήταν αναμενόμενο η τιμή του Kendall's $\tau$ αυξάνει με το $\beta$ για όλες τις τιμές που ελέγξαμε. Εντούτοις, όταν η τιμή του $\beta$ γίνεται 0.2 ή μεγαλύτερη, η ποιότητα των αρχικών συστάσεων αρχίζει να πέφτει, καθώς οι άμεσες σχέσεις που υπάρχουν μεταξύ των αντικειμένων αγνοούνται όλο και περισσότερο. Διαισθητικά, η κατάλληλη επιλογή των παραμέτρων εξαρτάται από το πόσο αραιός είναι ο χώρος των αντικειμένων. Τα πειράματά μας με το MovieLens dataset προτείνουν πως μια επιλογή του $\beta$ μεταξύ του 0.1 και του 0.15 και μια επιλογή του $\alpha$ μεταξύ του 0.8 και του 0.75, δίνουν πολύ καλά αποτελέσματα τόσο στην ποιότητα των συστάσεων όσο και στην ανθεκτικότητα απέναντι στην αραιότητα.

# 4. ΠΕΙΡΑΜΑΤΙΚΗ ΑΞΙΟΛΟΓΗΣΗ

# Κεφάλαιο 5

# Συμπεράσματα - Μελλοντική έρευνα

Στην παρούσα διπλωματική, παρουσιάσαμε ένα νέο πλαίσιο Συνεργατικής Διήθησης που φέρει το όνομα HIR και προτείνει αντικείμενα στους χρήστες ενός συστήματος, υπό μορφή λίστας. Η μέθοδός μας ακολουθεί μια γραφοθεωρητική προσέγγιση. Ο HIR μοντελοποιεί το πρόβλημα ως ένα γράφο του οποίου οι κόμβοι εκφράζουν τα αντικείμενα και οι ακμές αναπαριστούν τις αλληλεπιδράσεις ή τις ομοιότητες μεταξύ των αντικειμένων. Για τον υπολογισμό αυτών των αλληλεπιδράσεων βασιστήκαμε τόσο στην άμεση όσο και στην έμμεση συσχέτιση των αντικειμένων. Όσον αφορά την έμμεση συσχέτιση πατήσαμε στη διαίσθηση πίσω από τον *NCDawareRank* εκμεταλλευόμενοι την εγγενώς ιεραρχική δομή του χώρου αντικειμένων.

Ο αλγόριθμός μας αποδείχτηκε πως είναι πολύ αποδοτικός τόσο από υπολογιστικής άποψης όσο και από πλευράς αποθήκευσης. Το μητρώο άμεσης συσχέτισης είναι εγγενώς αραιό και προσαρμόζεται πολύ καλά στην αύξηση του αριθμού των χρηστών. Το μητρώο NCD εγγύτητας, παρουσιάζει πολύ μικρές απαιτήσεις αποθήκευσης, λόγω της παραγοντοποίησης που επιδέχεται.

Επιπρόσθετα, από υπολογιστικής άποψης, ο αλγόριθμός μας αποδεικνύεται πολύ αποδοτικός· ο πιο "βαρύς" υπολογισμός κάθε επανάληψης είναι ο Sparse Matrix-Vector (SpMxV) πολλαπλασιασμός $\boldsymbol{\pi}^\mathsf{T}\mathbf{C}$, ενώ ο αριθμός επαναλήψεων που απαιτείται για την εξαγωγή αποτελεσμάτων βέλτιστης ποιότητας, στα πειράματα αξιολόγησης που πραγματοποιήσαμε, ήταν πάντα πολύ μικρός. Συγκεκριμένα, εφαρμόσαμε τον HIR στο *movie recommendation πρόβλημα* με χρήστη του ευρέως γνωστού MovieLens dataset. Τα πειράματα έδειξαν, έπειτα από τη σύγκριση με τους άλλους state-of-the-art αλγορίθμους κατάταξης, πως ο HIR υπερτερεί στις ranking-based μετρικές απόδοσης, την micro-DOA και macro-DOA. Επιπλέον, ο αλγόριθμός μας αποδείχτηκε πως συμπεριφέρεται καλά ακόμα και αν ο χώρος των αντικειμένων είναι αραιός. Σε αυτή τη διαπίστωση οδηγηθήκαμε, έπειτα από τη διεξαγωγή πειραμάτων στα οποία προσομοιώσαμε το φαινόμενο τόσο της γενικής όσο και της

# 5. ΣΥΜΠΕΡΑΣΜΑΤΑ - ΜΕΛΛΟΝΤΙΚΗ ΕΡΕΥΝΑ

τοπικής αραιότητας.

Μια πολύ ενδιαφέρουσα ερευνητική κατεύθυνση που αξίζει να εξερευνηθεί, είναι το πως θα συμπεριφερθεί το σύστημά μας κατά την εισαγωγή περισσοτέρων από μια decompositions, οι οποίες βασίζονται σε πολλαπλά κριτήρια. Στη συγκεκριμένη εργασία, εξετάσαμε μόνο την περίπτωση που ο χώρος των ταινιών χωρίζεται σε κατηγορίες. Ωστόσο, σε ρεαλιστικές περιπτώσεις - όπου είναι διαθέσιμες πολλές πληροφορίες - μπορούν να υπάρξουν περισσότερες ομαδοποιήσεις των αντικειμένων.

Συμπερασματικά, ο HIR με την αξιοποίηση των NCD συνόλων εγγύτητας, ρίχνει νέο φως στο recommendation πρόβλημα και παρουσιάζει μια εύκολα γενικεύσιμη μέθοδο που παρέχει ένα διαισθητικά ελκυστικό και υπολογιστικά αποδοτικό θεωρητικό πλαίσιο, για την παραγωγή ποιοτικών συστάσεων.

# Βιβλιογραφία